\documentclass[
 amsmath,amssymb,
]{revtex4-1}
\usepackage{graphicx}
\usepackage{dcolumn}
\usepackage{bm}
\usepackage{color}
\usepackage{hyperref}
\usepackage{ulem}

\begin{document}

\title{Terahertz-optical intensity grating for creating high-charge, attosecond electron bunches}

\author{Jeremy Lim\footnote{Present address: Singapore University of Technology and Design, 8 Somapah road, Singapore 487372, Singapore}},
\affiliation{School of Physical and Mathematical Sciences Nanyang Technological University, Singapore 637371, Singapore}

\author{Yidong Chong}
\affiliation{School of Physical and Mathematical Sciences Nanyang Technological University, Singapore 637371, Singapore}
\affiliation{Centre for Disruptive Photonic Technologies, Nanyang Technological University, Singapore 637371, Singapore}

\author{Liang Jie Wong}
\email{wonglj@simtech.a-star.edu.sg}
\affiliation{Singapore Institute of Manufacturing Technology, Singapore 138634}
\vspace{10pt}
\date{\today}

\begin{abstract}
Ultrashort electron bunches are useful for applications like ultrafast imaging, coherent radiation production, and the design of compact electron accelerators. Currently, however, the shortest achievable bunches, at attosecond time scales, have only been realized in the single- or very few-electron regimes, limited by Coulomb repulsion and electron energy spread. Using \textit{ab initio} simulations and complementary theoretical analysis, we show that highly-charged bunches are achievable by subjecting relativistic (few MeV-scale) electrons to a superposition of terahertz and optical pulses. We provide two detailed examples that use realistic electron bunches and laser pulse parameters which are within the reach of current compact set-ups: one with bunches of $>$ 240 electrons contained within 20 as durations and 15 $\mathrm{\mu}$m radii, and one with final electron bunches of ~1 fC contained within sub-400 as durations and 8 $\mathrm{\mu}$m radii. Our results reveal a route to achieve such extreme combinations of high charge and attosecond pulse durations with existing technology.
\end{abstract}

%

%
\maketitle

\section{Introduction}
Electron bunches of femtosecond-to-attosecond-scale duration are useful tools for studying ultrafast atomic-scale processes, including structural phase transitions in condensed matter~\cite{Siwick2003An-Atomic-Level,Baum20074D-Visualizatio,Morrison2014A-photoinduced-,Gedik2007Nonequilibrium-,Musumec2010Capturing-ultra,ScianiMiller2011}, sub-cycle changes in oscillating electromagnetic waveforms~\cite{Ryabov2016Electron-micros}, and the dynamics of biological structures~\cite{Fitzpatrick2013Exceptional-rig,Anthony-W.-P.-Fitzpatrick20134D-Cryo-Electro}. High-density electron bunches of sub-femtosecond durations are potentially useful in high-resolution, time-resolved atomic diffraction~\cite{Baum2017NatPhys}, as sources of extreme-ultraviolet radiation through inverse Compton scattering~\cite{PhysRevLett.104.234801,PhysRevSTAB.14.070702,Kiefer_rel_elec_mirrors,PhysRevLett.119.254801}, and as injection bunches for compact charged-particle accelerators~\cite{RJEngland_DLA,FerrariE_ACHIP}. Existing schemes for electron bunch compression include the use of electrostatic elements~\cite{WangGedikIEEE2012}, time-varying fields within radio-frequency (RF) cavities~\cite{Gao12,vanOudheusden2010PRL,Chatelain2012,Kassier2012,Gliserin2015Sub-phonon-peri}, electromagnetic transients~\cite{Baum2007Attosecond-elec,Hilbert2009Temporal-lenses,WLJ2015NJP,PriebeNatPhoton,KozakNatPhys2017,PhysRevLett.120.103203,KealhoferSci2016,EhbergerOSA2017}, and a combination of optical laser pulses and dielectric membranes~\cite{Baum2017NatPhys}. In all of these schemes, space charge effects and velocity spread enforce a tradeoff between electron bunch charge and pulse duration. Consequently, whereas an electron bunch of pulse duration 0.1-1 ps may contain ~250 fC~\cite{vanOudheusden2010PRL}, electron bunches of attosecond-scale durations (attobunches) are typically realized with single or very few electrons~\cite{Baum2017NatPhys,PriebeNatPhoton,KozakNatPhys2017,PhysRevLett.120.103203,KealhoferSci2016,EhbergerOSA2017,Zewail187,Aidelsburger2010PNAS}.

Here, we use \textit{ab initio} numerical simulations and complementary analytical theory to show that high-charge electron bunches of attosecond-scale durations can be produced by interfering coherent terahertz and optical pulses. We study two regimes of operation: in the first regime, 5 MeV electrons are compressed into attobunches of about $20$ as duration, each containing $\sim 240$ electrons. In the second regime, 5 MeV electrons are compressed into bunches of $<400$ as duration, each containing $\sim 1$ fC of charge.  By comparison, theoretical predictions of electron bunch compression using realistic bunches have so far been limited to about 200 as~\cite{KozakNatPhys2017} in the single-electron regime. Experimentally, the shortest electron bunches produced to date lie in the single-electron regime, with durations of 655 as~\cite{PriebeNatPhoton} and 820 as~\cite{Baum2017NatPhys}, and indirect measurements indicating durations as short as 260 as~\cite{PhysRevLett.120.103203}. 

In addition, we obtain fully closed-form expressions for the dynamics of electrons subject to a general combination of counter-propagating pulses. Given a specific initial electron bunch configuration, these analytical tools enable us to predict various key features of our compression scheme, such as the bunch duration at focus (maximum compression), and the final kinetic energy (KE) spread. Our analytical predictions agree well with our \textit{ab initio} numerical simulation results in regimes where space charge effects are negligible. Our work complements existing theoretical formulations for the behavior of charged particles in counter-propagating electromagnetic fields, which have been confined to the sub-relativistic regime~\cite{Hilbert2009Temporal-lenses,PhysRevA.98.013407}.

\begin{centering}
\begin{figure}[ht!]
\centering
\includegraphics[width = 160mm]{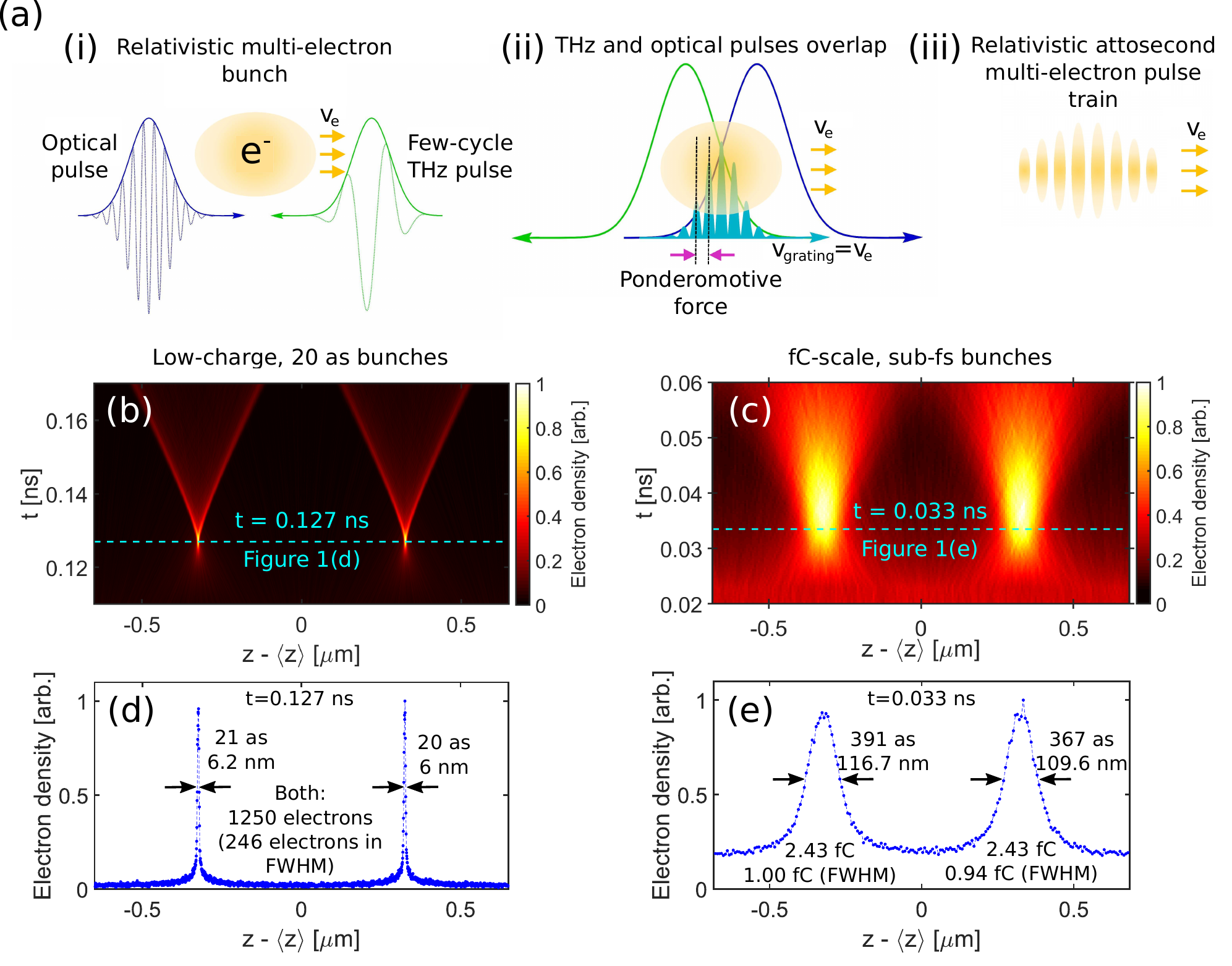}
\caption{High-charge, relativistic (5 MeV) attosecond electron pulses formed by a terahertz-optical intensity grating. The scheme we study is illustrated in (a): (i) A co-propagating optical pulse (blue waveform) and a counter-propagating few-cycle terahertz pulse (green waveform) are incident on a relativistic electron bunch (yellow ellipse) of mean velocity $v_{0}$. (ii) The pulses overlap, forming a sub-luminal intensity grating (solid turquoise profile) which co-propagates with the electron bunch. (iii) After the interaction, the impulse imparted by the grating compresses the electrons into a train of attobunches. The heatmaps in (b) and (c) show the electron density time-evolution using a centered coordinate system $z - \langle z \rangle$. The electron density spatial distributions at the focal times are shown in (d) and (e). In (b) and (d), electrons with $10^{-3} \%$ initial relative kinetic energy (KE) spread, and an average of 1250 electrons per grating period, interact with a 90.3 mJ optical pulse and a 39.0 $\mu$J terahertz pulse, resulting in bunches containing 246 electrons in $\sim20$ as durations (FWHM). In panels (c) and (e), a 20 fC bunch with a (FWHM) duration of about 16.5 fs and relative KE spread of 0.146\% interacts with a 6.66 mJ optical pulse and a 16.9 $\mu$J terahertz pulse, resulting in electron bunches of $< 400$ as (FWHM) containing approximately 1 fC of charge within the FWHM. Only the two central grating periods are plotted. The data in (b)-(e) are the result of electrodynamic simulations in which non-paraxial laser fields as well as near- and far-field space charge effects are exactly taken into account.}
\label{Fig_01}
\end{figure}
\end{centering}

In the proposed scheme, shown in figure~\ref{Fig_01}(a), the counter-propagating terahertz and optical pulses interfere to form an intensity grating, which is velocity-matched to the relativistic (few MeV-scale) electrons by choosing the proper carrier frequency for each pulse. The ponderomotive force, which is proportional to the negative intensity gradient, compresses the electrons into a train of attobunches. Bunch compression schemes based on intensity gratings have previously been studied only in the regime where both electromagnetic pulses are at optical/infrared frequencies for applications like electron acceleration~\cite{Hafizi1997Vacuum-beat-wav,Kozak2015Electron-accele,Esarey1995PRE}, and the compression of non-relativistic, single and few-electron bunches~\cite{Baum2007Attosecond-elec,Hilbert2009Temporal-lenses,KozakNatPhys2017,PhysRevLett.120.103203}. Here, we show that combining terahertz frequencies with optical frequencies creates an intensity grating that can be used to compress relativistic electron pulses achievable in lab-scale setups~\cite{Maxson2017Direct-Measurem,fs_time_res_diffraction,SLAC_MeV_rev_sci_instr,C4FD00204K} to attosecond scale durations with as much as 1 fC of charge per attobunch. We use counter-propagating terahertz pulses of $<100~\mathrm{\mu}$J and optical pulses of $<100$ mJ, which are readily obtained with today's technology~\cite{Yeh2007Generation-of-1,single_cycle_1THz,OR_organic_crystals,THz_DFG,HuangOptLett2013,Dhillon2017a,Fulop_THz_OR,Fulop_mJ_THz,THz_0.4mJ,THz_0.9mJ}. The absence of material structures in the interaction region of this scheme removes the possibility of material damage, allowing the intensity of our lasers to be scaled to arbitrarily high values for rapid focusing and strong compression of relativistic electron bunches. Due to the suppression of space charge effects at relativistic energies~\cite{HastingsApplPhysLett2006,MusumeciApplPhysLett2010}, the resulting attobunches can hold substantially higher charge than existing attobunches in the non-relativistic, single-electron regime~\cite{Baum2017NatPhys,KozakNatPhys2017,PhysRevLett.120.103203}.

Our \textit{ab initio} simulations (as described in the next section) exactly model the interactions of electrons with each other as well as with external laser fields. In particular, our simulations account for both near-field and far-field space charge effects, where near-field refers to fields responsible for the Coulomb force, and far-field refers to fields associated with radiation from the electron. We model the external laser fields using exact, finite-energy, non-paraxial solutions to Maxwell's equations. This is critical for accuracy since terahertz pulses from compact sources usually operate in the near-single-cycle limit and have beam waists tightly focused down to wavelength-scale dimensions~\cite{single_cycle_1THz} in order to achieve desired on-axis field strengths.

\section{Results}
\subsection{High-charge attosecond electron bunches}
Figure~\ref{Fig_01} presents results from 2 regimes of our study: (i) a regime where $\sim 20$ as electron bunch durations containing 246 electrons are realized and (ii) a regime where $<400$ as electron bunch durations are realized with fC-scale charge per bunch. The durations of the compressed bunches are stated using full width at half maximum (FWHM) values. In all simulation results presented in this section, the optical (co-propagating) and terahertz (counter-propagating) pulses have central wavelengths of $\lambda_{1} = 0.65~\mathrm{\mu m}$ and $\lambda_{2} = 300~\mathrm{\mu{m}}$ respectively, are linearly-polarized in $x$, and propagate in the $\pm z$ direction. The electron bunches have a mean KE of $\langle \mathrm{KE}\rangle =5$ MeV. The velocity of the intensity grating, $v_{gr}$, is matched to the mean velocity of the electrons, $v_{0}$, by choosing wavelengths, $\lambda_{1}$ and $\lambda_{2}$, such that~\cite{Baum2007Attosecond-elec,Esarey1995PRE}:
\begin{equation}
v_{gr} = v_{0} = c\bigg{(} \frac{\lambda_{2} - \lambda_{1}}{\lambda_{1}+\lambda_{2}} \bigg{)}
\label{eqn_standing_wave_condition}
\end{equation}
where $c$ is the speed of light in free space. In the lab frame, the grating period is given by
\begin{equation}
\lambda_{gr} = \frac{\lambda_{1}}{2\gamma_{0}^{2}(1 - \beta_{0})} = \frac{\lambda_{2}}{2\gamma_{0}^{2}(1 + \beta_{0})}
\end{equation}
where $\vec{\beta}=\beta_{0}\hat{z}=(v_{0}/c)\hat{z}$ is the normalized mean velocity of the electron bunch propagating in $\hat{z}$. The corresponding Lorentz factor is $\gamma_{0} = 1/\sqrt{1 - \beta_{0}^{2}}$. The mean KE of the electrons is $\langle \mathrm{KE} \rangle = (\gamma_{0} - 1)m_{e}c^{2}$, where $m_{e}$ is the electron rest mass.  (\ref{eqn_standing_wave_condition}) shows the necessity of combining very disparate counter-propagating laser wavelengths where relativistic electrons are concerned: for $v_{gr}$ close to the speed of light, $v_{0} \sim c$, $\lambda_{2} \gg \lambda_{1}$ is necessary. The use of relativistic electrons takes us into a regime beyond what has been studied for compressing non-relativistic electrons and gives us an opportunity to leverage the developments of high-intensity terahertz pulses in combination with optical pulses in our scheme.

Figures \ref{Fig_01}(b) and \ref{Fig_01}(d) show the electron density distribution obtained by averaging over 300 sets of \textit{ab initio} simulations using an initial 5 MeV electron bunch containing 2 fC of charge uniformly distributed across 10 grating periods. After the laser-electron interaction, each resulting attobunch has about 1250 electrons contained within each $\lambda_{gr}$, and 246 electrons within the FWHM duration of 20 as.  The non-paraxial optical and terahertz pulses have energies of 90.3 mJ and 39.0 $\mu$J respectively. The optical pulse has a duration of 80 fs (intensity FWHM) and a peak on-axis field strength $E_{01} \approx 4.96\times 10^{10}$ V/m. The terahertz pulse has a 1 ps duration (intensity FWHM) and a peak on-axis field strength $E_{02}\approx 2.95\times10^{8}$ V/m. Both laser pulses have the same waist radius, $w_{0}=450~\mathrm{\mu{m}}$. During interaction, the bunch has a radius of about $15~\mathrm{\mu{m}}$. The initial relative KE spread is $\sigma_{\mathrm{KE}}/\langle\mathrm{KE}\rangle = 10^{-3} \%$. While this value is small, relative KE spreads as low as $\sigma_{\mathrm{KE}}/\langle\mathrm{KE}\rangle = 4\times10^{-4} \%$ have been predicted for existing RF gun set-ups~\cite{PhysRevSTAB.18.120102}. The full set of electron bunch and laser pulse parameters, as well as a plot of the non-paraxial terahertz pulse electric field spatial profile is found in Supporting Information Section S.5(iv).

The second scenario, shown in figures \ref{Fig_01}(c) and \ref{Fig_01}(e), involves compressing an electron bunch of $\langle \mathrm{KE} \rangle=5$ MeV, 20 fC (total charge), 16.5 fs FWHM duration, relative energy spread $\sigma_{\mathrm{KE}}/\langle \mathrm{KE} \rangle\approx 0.146 \%$ and $8~\mathrm{\mu{m}}$ bunch radius, into a train of sub-400 as duration, fC-scale electron bunches. The electron density heatmap and distribution are averaged over 200 sets of \textit{ab initio} simulation results. The initial electron bunch was modelled after the bunch experimentally demonstrated in \cite{Maxson2017Direct-Measurem} (see Supporting Information  Section S.5(v)). Both pulsed lasers have the same beam waist: $w_{0} = 200~\mathrm{\mu{m}}$. The optical pulse has a duration of 30 fs (intensity FWHM) and an on-axis peak field strength $E_{01} \approx 5\times10^{10}$ V/m, corresponding to a pulse energy of 6.66 mJ. The terahertz pulse (figure S10 in Supporting Information Section S.5(v)) has a duration of 1 ps (intensity FWHM) and an on-axis peak field strength $E_{02}\approx4.18\times10^8$ V/m, corresponding to a pulse energy of 16.9 $\mu$J. Such optical and terahertz pulses are readily achievable today in a table-top setup~\cite{Yeh2007Generation-of-1,single_cycle_1THz,OR_organic_crystals,THz_DFG,HuangOptLett2013,Dhillon2017a,Fulop_THz_OR,Fulop_mJ_THz,THz_0.4mJ,THz_0.9mJ}.  At the focus, we observe the formation of electron bunches with about 1 fC of charge in a FWHM duration of 367 as (figure \ref{Fig_01}(e)). 

\begin{figure}[ht!]
\centering
\includegraphics[width = \textwidth]{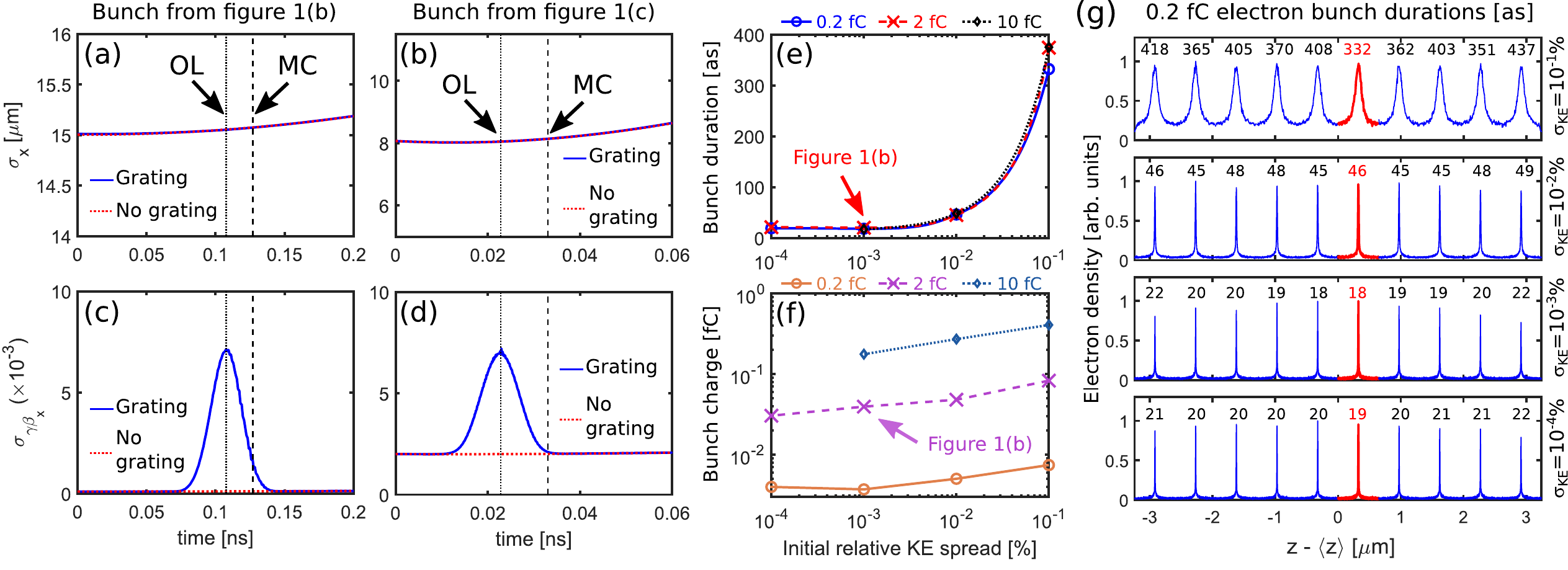}
\caption{Transverse dynamics of a 5 MeV multi-electron bunch and dependence of bunch duration (FWHM) and charge on initial KE spread. (a) and (c) show the evolving transverse dynamics of the bunch plotted in figures \ref{Fig_01}(b) and \ref{Fig_01}(d); (b) and (d) show the transverse dynamics of the bunch plotted in figures~\ref{Fig_01}(c) and \ref{Fig_01}(e). (a) and (b) show the evolution of the bunch radius, $\sigma_{x}$, while (c) and (d) show the evolution of the transverse normalized momentum, $\sigma_{\gamma\beta_{x}}$. The vertical dotted line labeled ``OL'' shows the time when the laser pulse peaks overlap and the vertical dashes labeled ``MC'' correspond to the time of maximum compression as shown in figure~\ref{Fig_01}. The cases with no electron-intensity grating interaction are plotted using the red dotted lines. The intensity grating imparts a significant momentum spread only during  interaction. (e) The bunch FWHM duration at the focus increases with increasing initial KE spread. Durations of about $40$ as can be achieved at spreads of $10^{-2}\%$, and durations of $\leq20$ as can be achieved for spreads of $10^{-3}\%$ and lower. (g) shows the electron density distribution at the time of maximum compression of the central well (red distribution), which are the values used to plot the 0.2 fC case in (e), for different values of initial electron KE spread. The bunch durations in attoseconds are indicated above each peak.}
\label{Fig_02}
\end{figure}

Figures \ref{Fig_02}(a) and \ref{Fig_02}(c) show the transverse dynamics induced by the intensity grating for the case studied in figures \ref{Fig_01}(b) and \ref{Fig_01}(d) while and those in figures \ref{Fig_02}(b) and \ref{Fig_02}(d) correspond to the case studied in figures \ref{Fig_01}(c) and \ref{Fig_01}(e).  The evolution of $\sigma_{x}$ and $\sigma_{\gamma\beta_{x}}$ with space-charge effects, but without laser-electron interaction, has been plotted using red dotted lines. It can be seen that the laser interaction imparts large transverse momenta in $x$ only during the time of interaction (vertical dotted line labeled ``OL''), but long after interaction, the transverse dynamics are practically indistinguishable from the case with no electron-intensity grating interaction.  For the case shown in figure \ref{Fig_02}(c), the compression is strong enough such that maximum compression (vertical dashes labeled ``MC'') occurs before the grating has completely faded. Thus, the transverse momentum spread is still significant ($\sigma_{\gamma\beta_{x}} = 1.28\times10^{-3}$) compared to the case without the grating ($\sigma_{\gamma\beta_{x}} = 0.11\times10^{-3}$). For the case shown in figure~\ref{Fig_02}(d), maximum compression is attained just after the intensity grating has faded. Hence, the transverse momentum spread at maximum compression ($\sigma_{\gamma\beta_{x}} = 2.11\times10^{-3}\%$) is similar to the case where there is no electron-grating interaction ($\sigma_{\gamma\beta_{x}} = 2.01\times10^{-3}$).  Hence, when low transverse momentum spread and bunch expansion is desired, care should be taken to ensure maximum compression is attained long after the intensity grating has faded.

Figures~\ref{Fig_02}(e) and \ref{Fig_02}(f) show the achievable electron bunch duration at the focus and the amount of charge contained within the FWHM duration as a function of initial electron KE spread for a fixed amount of total charge.  The laser pulse and electron bunch parameters used (except for the charge amount and initial KE spread) are the same as those used to produce figures~\ref{Fig_01}(b) and \ref{Fig_01}(d).  Figure~\ref{Fig_02}(e) indicates that with initial relative KE spreads on the order of $0.1\%$, which is achievable with the current state-of-the-art few-MeV beamlines~\cite{Maxson2017Direct-Measurem}, compressed bunches of durations on the order of hundreds of attoseconds can already be realized. For initial KE spreads on the order of about $10^{-2}\%$, sub-100 as bunches can be attained, and $\lesssim 10^{-3}\%$ initial KE spread yields bunches which have durations of 20 as and below at the focus. It should be noted that the charge contained within the attobunches can be enhanced by increasing the initial charge values without increasing the attobunch durations significantly (even up to 10 fC) due to the relativistic suppresion of space charge effects at few-MeV electron energies. Figure~\ref{Fig_02}(g) shows the electron density distribution for all attobunches at the time of maximum compression of the attobunch closest to the grating center (red distribution, values used to plot figure~\ref{Fig_02}(e)) for the 0.2 fC case.  Our results indicate that despite each attobunch having differing focal times which depend on their relative distance from the center of the intensity grating, the final bunch durations across the entire macrobunch are similar, and by appropriate selection of laser pulse durations, the focal times for each attobunch can be controlled (Supporting Information Section S.2).  Our results show that a combination of terahertz and optical technologies can be enabling concepts for the realization of high-charge electron bunches of sub-fs durations.

\subsection{Theoretical predictions of key bunch parameters}
\begin{figure}[ht!]
\centering
\includegraphics[width = \textwidth]{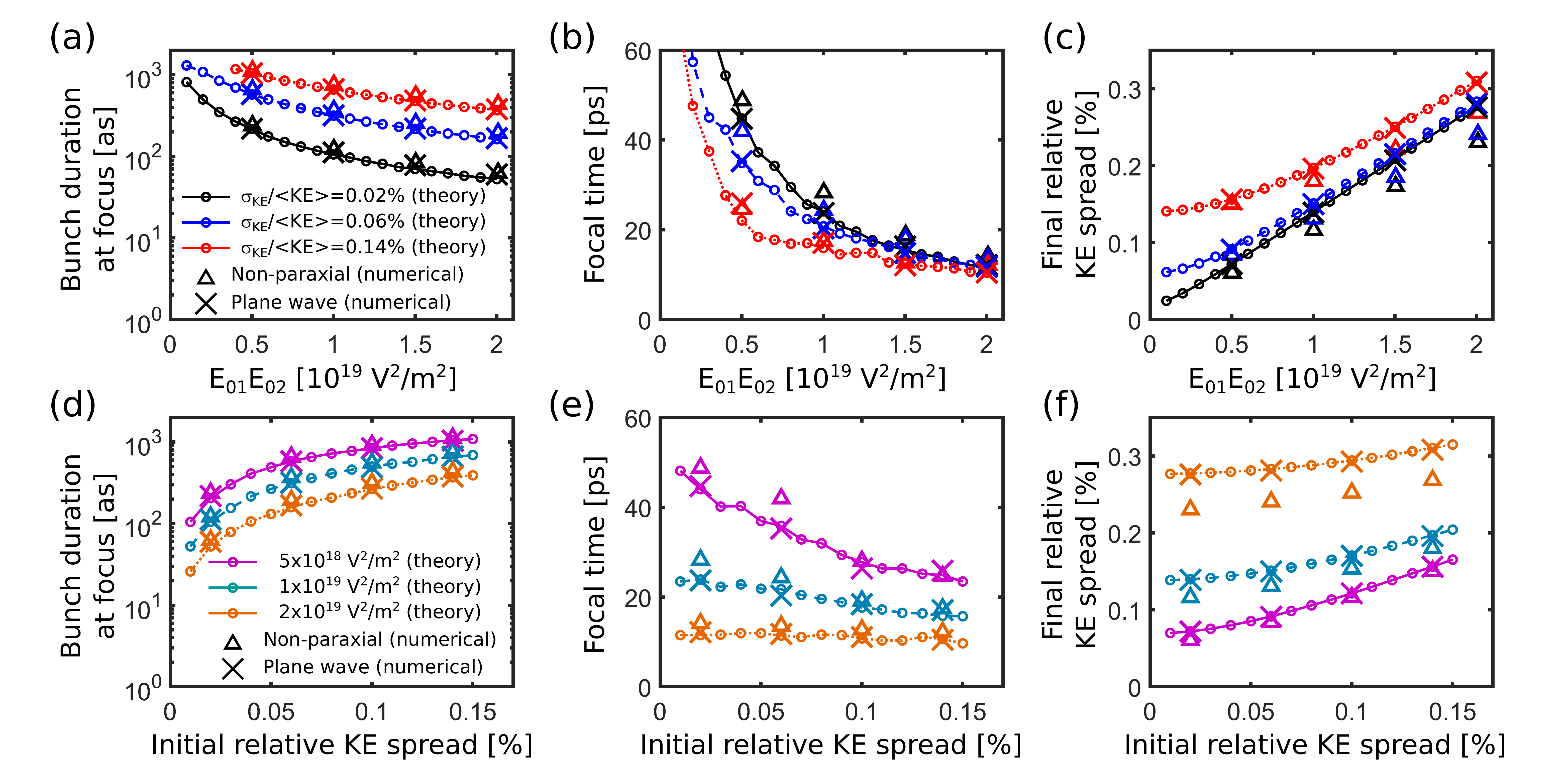}
\caption{Dependence of compressed electron bunch properties on laser field amplitudes and initial relative KE spread.  $E_{01}E_{02}$ denotes product of the peak field strengths (on-axis values for non-paraxial case), which is varied as a parameter in (a)-(c).  $\sigma_{\mathrm{KE}}/\langle \mathrm{KE}\rangle$ is the initial relative KE spread and is varied as a parameter in (d)-(f). The FWHM duration at focus is plotted in (a) and (d), the focal time in (b) and (e), and the final KE spread in (c) and (f). Circles indicate theory, crosses indicate simulations where the laser fields are modelled as plane wave pulses, and the triangles indicate simulations where the laser fields are modelled as exact non-paraxial pulses ($w_{0} = 300~\mathrm{\mu{m}}$). The terahertz and optical pulses have FWHM durations of 1 ps and 30 fs respectively. The electron bunch mean KE is 5 MeV and has a radius of 10 $\mu$m. Space charge effects are neglected in this comparison. We note the excellent agreement without the theoretical predictions and plane wave simulation results. Discrepancies between plane wave and non-paraxial simulations results show the non-trivial influence of the transverse laser pulse profile in our scenarios.}
\label{Fig_03}
\end{figure}

We now present fully closed-form expressions for the behavior of charged particles subject to a pair of counter-propagating electromagnetic pulses. These expressions, which neglect space charge effects, have been used to predict various key properties of our bunch compression scheme -- including the focal time (maximum compression), the bunch duration at focus, and the final KE spread -- and show excellent agreement with the results of our \textit{ab initio} simulations in regimes where space charge and non-paraxial laser pulse effects are small (see figure \ref{Fig_03}).

We start from the Newton-Lorentz equations of motion, which describe the dynamics of electrons in arbitrary electromagnetic fields. Treating the counter-propagating laser pulses as pulsed plane waves and considering an electron moving in an arbitrary direction such that the transverse ($x,y$-direction) momenta are small compared to the longitudinal ($z$-direction) momentum, we obtain the normalized electron velocity long after interaction as (see Supporting Information  Sections S.1 to S.3 for detailed derivations):
\begin{equation}
\beta_{z,f}' \approx \Bigg{(} \sqrt{\frac{\pi}{\alpha_{a}}}\frac{T_{1}'}{\omega'}\cos(\Delta\theta)\frac{e^{2}E_{01}'E_{02}'}{m_{e}^{2}c^{2}}\sin(2k'z_{OLe}' + \phi_{0})\exp\Bigg{\{} \frac{-4(z_{OLe}' - z_{OL}')^{2}}{c^{2}[T_{1}'^{2}(1+\beta_{z,i}')^{2} + T_{2}'^{2}(1 - \beta_{z,i}')^{2}]} \Bigg{\}} \Bigg{)} + \beta_{z,i}'
\label{eqn_betafp}
\end{equation}
where the primes on the variables indicate that they are evaluated in the frame moving at normalized velocity $\vec{\beta}=\beta_{0}\hat{z}$. We define this to be the primed frame. For our electron bunch compression scheme, we take $\beta_{0}$ as the mean normalized velocity of the electron bunch being compressed. $E_{0j}'$ and $T_{j}'$ respectively refer to the electric field amplitude and pulse duration of the laser pulse labelled by subscript $j$, where $j=1$ ($j=2$) refers to the laser pulse which co-propagates (counter-propagates) with respect to the electron bunch. $\omega' = k'c$ is the central angular frequency of the laser pulses (which have the same frequency in the primed frame), $\Delta\theta$ is the relative angle between the polarization vectors associated with the two laser pulses (which we set to $0$ here for the strongest compression), $\phi_{0}$ is a phase constant that depends on the carrier envelope phase of each laser pulse, and $\beta_{z,i}'$ is the initial normalized electron speed. The intensity peaks of the counter-propagating laser pulses overlap at position $z' = z_{OL}'$ and time $t' = t_{OL}'$, and we define the longitudinal electron position at the time $t' = t_{OL}'$ to be $z_{OLe}'$ in the limit where the laser field strengths go to zero. $\alpha_{a}$ is defined as
\begin{equation}
\alpha_{a} \equiv (1 - \beta_{z,i}')^{2} + \frac{T_{1}'^{2}}{T_{2}'^{2}}(1+\beta_{z,i}')^{2}.
\end{equation}
We also obtain the corresponding electron position long after interaction as
\begin{equation}
\begin{split}
z_{f}'(t') &= \beta_{z,f}'ct' + z_{OLe}' - \beta_{z,i}'ct_{OL}'\\
&+\Bigg{(} \frac{\alpha_{b}}{\alpha_{a}}\sqrt{\frac{\pi}{\alpha_{a}}}\frac{T_{1}'}{2\omega'}\cos(\Delta\theta)\frac{e^{2}E_{01}'E_{02}'}{m_{e}^{2}c}\sin(2k'z_{OLe}' + \phi_{0})
\exp\Bigg{\{} \frac{-4(z_{OLe}' - z_{OL}')^{2}}{c^{2}[T_{1}'^{2}(1 + \beta_{z,i}')^{2} + T_{2}'^{2}(1 - \beta_{z,i}')^{2}]} \Bigg{\}}  
\Bigg{)}
\end{split}
\label{eqn_zfp}
\end{equation}
where 
\begin{equation}
\alpha_{b} = \frac{2}{c}\bigg{\{} \frac{T_{1}'^{2}}{T_{2}'^{2}}(1+\beta_{z,i}')[z_{OLe}' - z_{OL}' - (1+\beta_{z,i}')ct_{OL}'] -(1 - \beta_{z,i}')[z_{OLe}' - z_{OL}' + (1 - \beta_{z,i}')ct_{OL}'] \bigg{\}}.
\end{equation}
When the bunch has vanishing longitudinal velocity spread, i.e. $\beta_{z,i}'=0$, the general expression for the focal time, defined as the time between $t_{OL}'$ and the electrons reaching maximum compression, $t_{comp}'$ , is:
\begin{equation}
t_{comp}' - t_{OL}' = \frac{m_{e}}{K_{0}'\sqrt{\pi}}\sqrt{\frac{1}{T_{1}'^{2}} + \frac{1}{T_{2}'^{2}}}\exp\Bigg{[} \frac{4(z_{OLe}' - z_{OL}')^{2}}{c^{2}(T_{1}'^{2} + T_{2}'^{2})} \Bigg{]} + \frac{z_{OLe}' - z_{OL}'}{c}\Bigg{(} \frac{T_{2}'^{2} - T_{1}'^{2}}{T_{1}'^{2}+T_{2}'^{2}} \Bigg{)}.
\label{eqn_foc_times}
\end{equation}
Here, $K_{0}' = (2e^{2}E_{01}'E_{02}'\cos\Delta\theta)/(m_{e}c^{2})$.  In the special case where we consider the electrons near the center of the intensity grating ($z_{OLe}' \approx z_{OL}'$) and $T_{1}' = T_{2}' = T'$,$E_{01}' = E_{02}' = E_{0}'$ , (\ref{eqn_foc_times}) reduces to
\begin{equation}
t_{comp}' - t_{OL}' = \Bigg{(} \frac{m_{e}}{K_{0}'}\sqrt{\frac{2}{\pi}}\Bigg{)}\frac{1}{T'}
\end{equation}
which agrees with the analytical result obtained in \cite{Hilbert2009Temporal-lenses}, modulo a factor of $\sqrt{2/\pi}$ which comes from our choice of a Gaussian pulsed profile.

The overlap of the optical and terahertz pulses results in a finite-length intensity grating in which electrons farther from the center of the intensity grating generally experience a weaker compressive force. This effect is taken into account through the exponential factors in (\ref{eqn_betafp}) and (\ref{eqn_zfp}), as well as through $\alpha_{b}$.

The results in figure \ref{Fig_03} show the excellent agreement between our analytical predictions (circles) and numerical results when the laser pulses are modelled as pulsed plane waves (crosses). The discrepancy between the plane wave simulations and the exact numerical results using non-paraxial pulses (triangles) shows the importance of taking into account the transverse profiles of the focused optical and terahertz pulses in our simulations. Nevertheless, we also note that these exact results follow the trend predicted by our theory relatively well in the regime considered in figure \ref{Fig_03}.  In figure \ref{Fig_03}, the 5 MeV, $10~\mathrm{\mu{m}}$-radius electron bunch was modelled using $3.75\times10^{5}$ particles, and has a uniform random distribution in $z$ over a length of $\lambda_{gr}$. The initial bunch is normally-distributed in $x$ and $y$. The initial momentum spread for all cases is normally-distributed in all directions and isotropic: $\sigma_{\gamma\beta_{x}} = \sigma_{\gamma\beta_{y}} = \sigma_{\gamma\beta_{z}}$. We used the following initial relative KE spreads: $\sigma_{\mathrm{KE}}/\langle\mathrm{KE}\rangle= $ 0.02\%, 0.06\%, 0.10\%, and 0.14\%. The corresponding momentum spreads are $\sigma_{\gamma\beta_{i}}=1.9615\times10^{-3}$, $5.8848\times10^{-3}$, $9.8075\times10^{-3}$, and $1.3731\times10^{-2}$ respectively ($i\in\{x,y,z\}$). All electron bunch and laser pulse parameters are listed in Supporting Information  Section S.5(vi). 

Figures \ref{Fig_03}(a) and \ref{Fig_03}(d) show that a larger initial electron bunch KE spread makes it more difficult to compress the bunch unless higher laser field strengths field strengths are used. Figure \ref{Fig_03} thus highlights the importance of low energy spread in realizing attosecond bunches. As seen in figure \ref{Fig_03}(a), a change in relative initial KE spread from 0.02\% to 0.14\% can cause the electron bunch durations at the focus to increase by almost an order of magnitude. In the limit where $E_{01}E_{02}$ is small, we see from figure \ref{Fig_03}(a)-(c) that it is possible to obtain fs-scale electron bunches with a very small (practically negligible) change in energy spread, at the cost of a longer focal time. In figure \ref{Fig_03}(b), the decrease in the focal time approximately as $1/E_{01}'E_{02}' = 1/E_{01}E_{02}$ agrees with the trend predicted by (\ref{eqn_foc_times}).

Using the formalism described here, the predicted durations and focal times for the cases shown in figures~\ref{Fig_01}(b)-\ref{Fig_01}(e) are also in good agreement with our \textit{ab initio} simulations. For the case shown in figures~\ref{Fig_01}(b) and \ref{Fig_01}(d), the predicted FWHM duration for both the left and right attobunches is $9$ as, which is a good estimate of the numerically computed values of $21$ as and $20$ as; the theoretical time of maximum compression is $0.123$ ns, which is very close to the actual value of $0.127$ ns.  For the case shown in figures~\ref{Fig_01}(c) and \ref{Fig_01}(e), the theoretically predicted durations of the left and right attobunches are $352$ as and $338$ as respectively while the numerically computed durations are $391$ as and $367$. The theoretical time of maximum compression is $0.032$ ns, which is very close to the actual value of $0.033$ ns.

\section{Discussion}
\begin{figure}[ht!]
\centering
\includegraphics[width = 100mm]{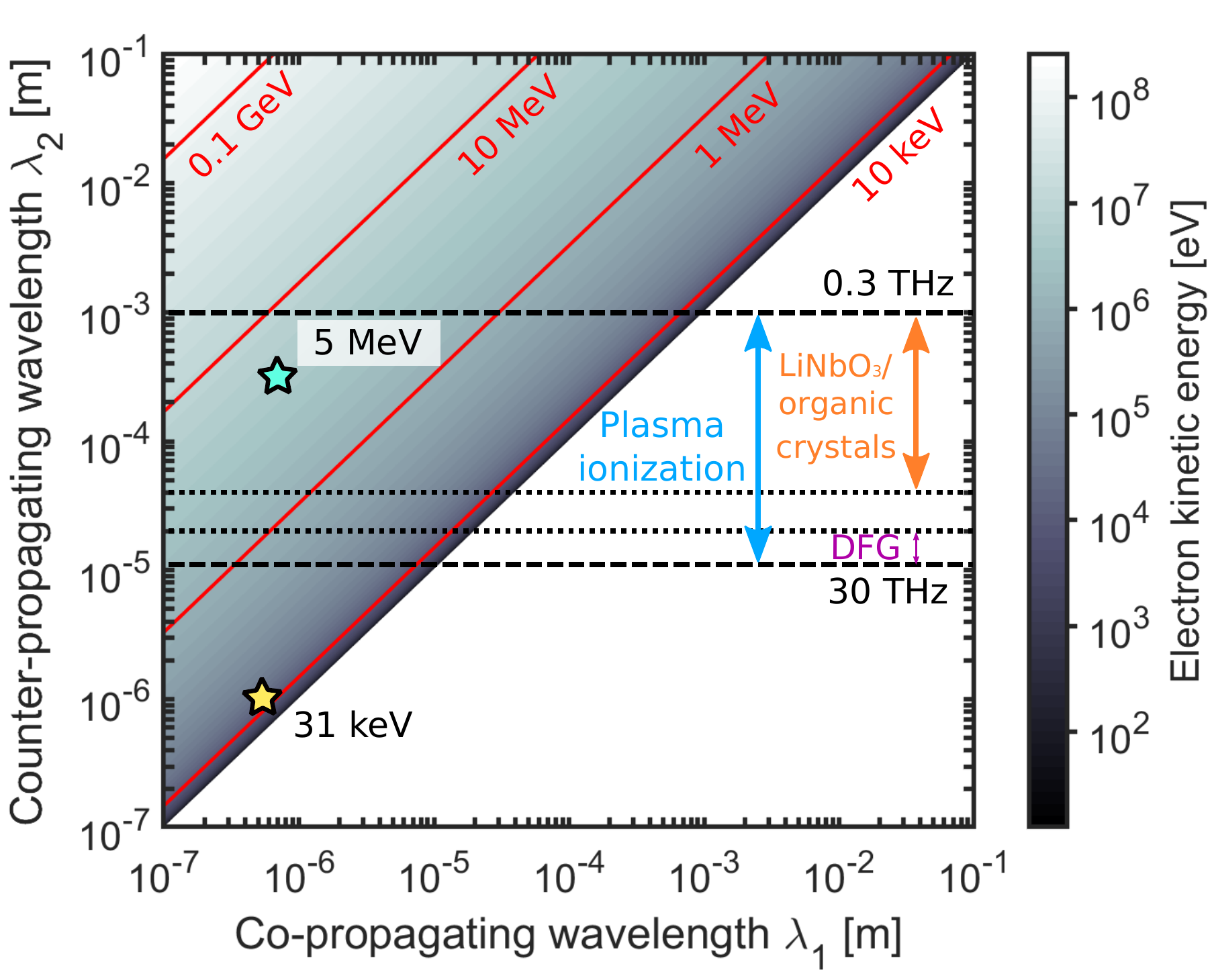}
\caption{Operating regimes for electron pulse compression. The colormap shows the electron bunch kinetic energies which can be matched by a range of co-propagating and counter-propagating wavelengths ($\lambda_{1}$ and $\lambda_{2}$ respectively). Only the region corresponding to $\lambda_{1} < \lambda_{2}$ is plotted. The region in which $\lambda_{1} > \lambda_{2}$ corresponds to a counter-propagating grating. The region bounded by the black dashed lines correspond to the terahertz regime~\cite{Dhillon2017a} for $\lambda_{2}$, and the black dotted lines within further divide this frequency range into bandwidths that are currently attainable through optical rectification of LiNbO3 and organic crystals, difference frequency generation (DFG), and plasma ionization. The yellow star marks the 31 keV, non-relativistic case, studied in~\cite{Baum2007Attosecond-elec}, whereas the blue star marks the 5 MeV, relativistic case which we study in this paper.}
\label{Fig_04}
\end{figure}

Here, we present an overview of the electron kinetic energies which can be matched using sources of coherent light at various wavelengths, as well as a brief comparison between our scheme and existing electron bunch compression schemes. The interest in working with electrons of larger kinetic energies is due to the relativistic suppression of space charge effects, which allows shorter bunch durations to be achieved in this compression scheme. The development of intense, coherent terahertz sources on a table-top scale~\cite{Yeh2007Generation-of-1,single_cycle_1THz,OR_organic_crystals,THz_DFG,HuangOptLett2013,Dhillon2017a,Fulop_THz_OR,Fulop_mJ_THz,THz_0.4mJ,THz_0.9mJ} as figure \ref{Fig_04} shows, unlocks a range of electron kinetic energies spanning 4 orders of magnitude (keV to 10 MeV). By contrast, using only wavelengths falling in the optical to near-infrared regime (0.4 $\mathrm{\mu{m}}$ to 1.4 $\mathrm{\mu{m}}$) would limit us to electron kinetic energies of 100 keV or less. 

Although the mechanism here can be extended to electron kinetic energies on the order of $10^{2}$ MeV and higher, much larger laser intensities would be involved for effective compression. The study of the use of this mechanism for such ultrarelativistic electrons is beyond the scope of this work. We note that alternative techniques for producing highly-compressed electron bunches of kinetic energies from tens-of-MeV to GeV include the use of compact inverse free electron laser systems~\cite{PRL_hightrapping_efficiency} and undulator modulators~\cite{PRL_single_cycle_XUV} have already been demonstrated and proposed. These methods may be more practical when larger dedicated accelerator facilities are available. However, for compact acceleration schemes such as dielectric laser acceleration~\cite{RJEngland_DLA}, the ability to produce fC-scale, few-MeV electron bunches modulated to sub-fs scales as injection sources, like those presented in this work, are of interest.

A number of laser-based sources of intense terahertz radiation, suitable for the use in the present compression scheme, as well as other forms of charged particle manipulation, have already been reported in the literature. Single-cycle and quasi-single-cycle terahertz radiation centered at 1 to 2 THz with peak field strengths on the order of 1 MV/cm have been achieved using optical rectification of LiNbO3 with tilted pulse front pumping~\cite{Yeh2007Generation-of-1,single_cycle_1THz} and optical rectification of organic crystals with high non-linear constants~\cite{OR_organic_crystals}. Semiconductors have also been shown to be a promising alternative for generating high-energy THz pulses using this technique~\cite{Fulop_THz_OR}. Terahertz pulse energies on the order of tens of $\mu$J are already routinely produced~\cite{Dhillon2017a} from these compact sources and energies up to 1 mJ~\cite{THz_0.9mJ,THz_0.4mJ} have already been demonstrated. With the high field strengths accompanying these high pulse energies, shorter attobunch durations and focusing times can be achieved, which our results in figure~\ref{Fig_03} predicts. The development of compact THz sources of higher energies, would alleviate the need for extremely tight-focusing of the THz pulse in order to achieve the desired field strengths.

Difference frequency generation (DFG) of optical parameteric amplifiers have been used to produced narrow-band, multi-cycle pulses at mid-infrared frequencies (15-30 terahertz) and higher fields strengths of 100 MV/cm~\cite{THz_DFG}. While ultra-broadband terahertz radiation can be produced using plasma ionization~\cite{THz_plasma_ionization}, the field strengths are typically lower than those achieved using optical rectification. However, they could potentially be used for the compression of low-charge or single-electron bunches with small energy spreads over longer focal distances.

We note that greater flexibility in our choice of wavelength for matching a given electron kinetic energy can be achieved by tilting the counter-propagating pulses~\cite{Hilbert2009Temporal-lenses,KozakNatPhys2017,PhysRevLett.120.103203,Kozak2015Electron-accele}. In this case, however, too large a tilt angle will lead to restrictions on the transverse size of the electron bunch. Nevertheless, the concept of tilting laser pulses could be implemented in the terahertz-optical scheme to accommodate an even wider range of electron kinetic energies.

Dielectric membranes, in combination with an optical laser pulse, have been used to compress non-relativistic (70 keV), single-electron bunches to attosecond-scale durations~\cite{Baum2017NatPhys}. When non-relativistic electrons are considered, the laser field strength required to modulate the bunch remains low enough to avoid material damage. However, relativistic electron bunches require much higher intensities for compression to attosecond time-scales, making material damage more likely. The scheme studied in the present paper allows high-intensity lasers to be used without the risk of material damage.

\section{Conclusion}
We presented a scheme in which counter-propagating terahertz and optical pulses are used to compress relativistic electrons into a train of attosecond-duration bunches. Due to the space-charge suppression at few MeV-scale energies, significant amounts of charge can be contained within each attobunch, compared to previously realized attobunches that have only single or very few electrons. Our \textit{ab initio} simulations take near- and far-field space charge effects (associated with the Coulomb force and the electron radiation respectively) into account, and use exact, non-paraxial pulse profiles to model single-cycle, tightly-focused terahertz pulses; this is a significant advance over previous numerical studies of similar intensity grating compression schemes, which assumed non-interacting electrons and planar or paraxial electromagnetic waves.

We presented results for attosecond electron bunch compression in two regimes. The first case involved the compression of a lower-charge electron cloud into attobunches with durations of about $20$  (FWHM), containing about 246 electrons. Such short-duration bunches could be used, for instance, as sources of high-quality coherent radiation through processes like inverse Compton scattering~\cite{Kiefer_rel_elec_mirrors}, Smith-Purcell radiation~\cite{Sergeeva2017Smith-Purcell-r}, transition radiation~\cite{Zhang2017Transition-radi}, and through electron-plasmon scattering~\cite{WLJ_nat_photon_2016,Rosolen_LightSci_2018}. We find that the realization of this scenario depends on having kinetic energy spreads which are extremely low but feasible~\cite{PhysRevSTAB.18.120102}.  In the second, the initial electron bunch contains 20 fC of charge and is comparable to the bunches that can be produced by existing few-MeV scale electron sources. In this case, we showed that the electrons can be compressed into smaller bunches of sub-400 as durations (FWHM), each containing up to 1 fC of charge. Besides electron diffraction applications (e.g. time-resolved atomic diffraction in~\cite{Baum2017NatPhys}), these bunches could potentially serve as pre-accelerated injection sources for compact dielectric laser acceleration (DLA) schemes, in which fC-scale, few-MeV electron bunches are desirable as input~\cite{RJEngland_DLA}. The modulated sub-fs bunches generated by our scheme can fit into the phase space acceleration buckets -- typically also of sub-laser wavelength length-scales -- which could improve the accelerated beam quality~\cite{RJEngland_DLA}. The sub-micron transverse bunch dimensions required for injection into typical optical DLA schemes can be achieved through the use of electron beam focusing optics.  Our results indicate that attosecond-scale electron bunches are not inherently limited to the few-to-single-electron regime, which has been the focus of other studies.

\begin{acknowledgements}
We thank the National Supercomputing Center (NSCC) Singapore for the use of their computing resources. LJW acknowledges support from the Science and Engineering Research Council (SERC; grant no. 1426500054) of the Agency for Science, Technology and Research (A*STAR), Singapore.  All authors made critical contributions to this manuscript and declare no competing financial interests.  
\end{acknowledgements}


\begin{thebibliography}{60}%
\makeatletter
\providecommand \@ifxundefined [1]{%
 \@ifx{#1\undefined}
}%
\providecommand \@ifnum [1]{%
 \ifnum #1\expandafter \@firstoftwo
 \else \expandafter \@secondoftwo
 \fi
}%
\providecommand \@ifx [1]{%
 \ifx #1\expandafter \@firstoftwo
 \else \expandafter \@secondoftwo
 \fi
}%
\providecommand \natexlab [1]{#1}%
\providecommand \enquote  [1]{``#1''}%
\providecommand \bibnamefont  [1]{#1}%
\providecommand \bibfnamefont [1]{#1}%
\providecommand \citenamefont [1]{#1}%
\providecommand \href@noop [0]{\@secondoftwo}%
\providecommand \href [0]{\begingroup \@sanitize@url \@href}%
\providecommand \@href[1]{\@@startlink{#1}\@@href}%
\providecommand \@@href[1]{\endgroup#1\@@endlink}%
\providecommand \@sanitize@url [0]{\catcode `\\12\catcode `\$12\catcode
  `\&12\catcode `\#12\catcode `\^12\catcode `\_12\catcode `\%12\relax}%
\providecommand \@@startlink[1]{}%
\providecommand \@@endlink[0]{}%
\providecommand \url  [0]{\begingroup\@sanitize@url \@url }%
\providecommand \@url [1]{\endgroup\@href {#1}{\urlprefix }}%
\providecommand \urlprefix  [0]{URL }%
\providecommand \Eprint [0]{\href }%
\providecommand \doibase [0]{http://dx.doi.org/}%
\providecommand \selectlanguage [0]{\@gobble}%
\providecommand \bibinfo  [0]{\@secondoftwo}%
\providecommand \bibfield  [0]{\@secondoftwo}%
\providecommand \translation [1]{[#1]}%
\providecommand \BibitemOpen [0]{}%
\providecommand \bibitemStop [0]{}%
\providecommand \bibitemNoStop [0]{.\EOS\space}%
\providecommand \EOS [0]{\spacefactor3000\relax}%
\providecommand \BibitemShut  [1]{\csname bibitem#1\endcsname}%
\let\auto@bib@innerbib\@empty
\bibitem [{\citenamefont {Siwick}\ \emph {et~al.}(2003)\citenamefont {Siwick},
  \citenamefont {Dwyer}, \citenamefont {Jordan},\ and\ \citenamefont
  {Miller}}]{Siwick2003An-Atomic-Level}%
  \BibitemOpen
  \bibfield  {author} {\bibinfo {author} {\bibfnamefont {B.~J.}\ \bibnamefont
  {Siwick}}, \bibinfo {author} {\bibfnamefont {J.~R.}\ \bibnamefont {Dwyer}},
  \bibinfo {author} {\bibfnamefont {R.~E.}\ \bibnamefont {Jordan}}, \ and\
  \bibinfo {author} {\bibfnamefont {R.~J.~D.}\ \bibnamefont {Miller}},\
  }\href@noop {} {\bibfield  {journal} {\bibinfo  {journal} {Science}\ }\textbf
  {\bibinfo {volume} {302}},\ \bibinfo {pages} {1382 } (\bibinfo {year}
  {2003})}\BibitemShut {NoStop}%
\bibitem [{\citenamefont {Baum}\ \emph {et~al.}(2007)\citenamefont {Baum},
  \citenamefont {Yang},\ and\ \citenamefont
  {Zewail}}]{Baum20074D-Visualizatio}%
  \BibitemOpen
  \bibfield  {author} {\bibinfo {author} {\bibfnamefont {P.}~\bibnamefont
  {Baum}}, \bibinfo {author} {\bibfnamefont {D.~S.}\ \bibnamefont {Yang}}, \
  and\ \bibinfo {author} {\bibfnamefont {A.~H.}\ \bibnamefont {Zewail}},\
  }\href@noop {} {\bibfield  {journal} {\bibinfo  {journal} {Science}\ }\textbf
  {\bibinfo {volume} {318}},\ \bibinfo {pages} {788} (\bibinfo {year}
  {2007})}\BibitemShut {NoStop}%
\bibitem [{\citenamefont {Morrison}\ \emph {et~al.}(2014)\citenamefont
  {Morrison} \emph {et~al.}}]{Morrison2014A-photoinduced-}%
  \BibitemOpen
  \bibfield  {author} {\bibinfo {author} {\bibfnamefont {V.~R.}\ \bibnamefont
  {Morrison}} \emph {et~al.},\ }\href@noop {} {\bibfield  {journal} {\bibinfo
  {journal} {Science}\ }\textbf {\bibinfo {volume} {346}},\ \bibinfo {pages}
  {445} (\bibinfo {year} {2014})}\BibitemShut {NoStop}%
\bibitem [{\citenamefont {Gedik}\ \emph {et~al.}(2007)\citenamefont {Gedik},
  \citenamefont {Yang}, \citenamefont {Logvenov}, \citenamefont {Bozovic},\
  and\ \citenamefont {Zewail}}]{Gedik2007Nonequilibrium-}%
  \BibitemOpen
  \bibfield  {author} {\bibinfo {author} {\bibfnamefont {N.}~\bibnamefont
  {Gedik}}, \bibinfo {author} {\bibfnamefont {D.~S.}\ \bibnamefont {Yang}},
  \bibinfo {author} {\bibfnamefont {G.}~\bibnamefont {Logvenov}}, \bibinfo
  {author} {\bibfnamefont {I.}~\bibnamefont {Bozovic}}, \ and\ \bibinfo
  {author} {\bibfnamefont {A.~H.}\ \bibnamefont {Zewail}},\ }\href@noop {}
  {\bibfield  {journal} {\bibinfo  {journal} {Science}\ }\textbf {\bibinfo
  {volume} {316}},\ \bibinfo {pages} {425} (\bibinfo {year}
  {2007})}\BibitemShut {NoStop}%
\bibitem [{\citenamefont {Musumeci}\ \emph
  {et~al.}(2010{\natexlab{a}})\citenamefont {Musumeci}, \citenamefont {Moody},
  \citenamefont {Scoby}, \citenamefont {Gutierrez}, \citenamefont {Westfall},\
  and\ \citenamefont {Li}}]{Musumec2010Capturing-ultra}%
  \BibitemOpen
  \bibfield  {author} {\bibinfo {author} {\bibfnamefont {P.}~\bibnamefont
  {Musumeci}}, \bibinfo {author} {\bibfnamefont {J.~T.}\ \bibnamefont {Moody}},
  \bibinfo {author} {\bibfnamefont {C.~M.}\ \bibnamefont {Scoby}}, \bibinfo
  {author} {\bibfnamefont {M.~S.}\ \bibnamefont {Gutierrez}}, \bibinfo {author}
  {\bibfnamefont {M.}~\bibnamefont {Westfall}}, \ and\ \bibinfo {author}
  {\bibfnamefont {R.~K.}\ \bibnamefont {Li}},\ }\href {\doibase
  10.1063/1.3520283} {\bibfield  {journal} {\bibinfo  {journal} {J. Appl.
  Phys.}\ }\textbf {\bibinfo {volume} {108}},\ \bibinfo {pages} {114513}
  (\bibinfo {year} {2010}{\natexlab{a}})}\BibitemShut {NoStop}%
\bibitem [{\citenamefont {Sciaini}\ and\ \citenamefont
  {Miller}(2011)}]{ScianiMiller2011}%
  \BibitemOpen
  \bibfield  {author} {\bibinfo {author} {\bibfnamefont {G.}~\bibnamefont
  {Sciaini}}\ and\ \bibinfo {author} {\bibfnamefont {R.~J.~D.}\ \bibnamefont
  {Miller}},\ }\href {http://stacks.iop.org/0034-4885/74/i=9/a=096101}
  {\bibfield  {journal} {\bibinfo  {journal} {Rep. Prog. Phys.}\ }\textbf
  {\bibinfo {volume} {74}},\ \bibinfo {pages} {096101} (\bibinfo {year}
  {2011})}\BibitemShut {NoStop}%
\bibitem [{\citenamefont {Ryabov}\ and\ \citenamefont
  {Baum}(2016)}]{Ryabov2016Electron-micros}%
  \BibitemOpen
  \bibfield  {author} {\bibinfo {author} {\bibfnamefont {A.}~\bibnamefont
  {Ryabov}}\ and\ \bibinfo {author} {\bibfnamefont {P.}~\bibnamefont {Baum}},\
  }\href@noop {} {\bibfield  {journal} {\bibinfo  {journal} {Science}\ }\textbf
  {\bibinfo {volume} {353}},\ \bibinfo {pages} {374} (\bibinfo {year}
  {2016})}\BibitemShut {NoStop}%
\bibitem [{\citenamefont {Fitzpatrick}\ \emph
  {et~al.}(2013{\natexlab{a}})\citenamefont {Fitzpatrick}, \citenamefont
  {Park},\ and\ \citenamefont {Zewail}}]{Fitzpatrick2013Exceptional-rig}%
  \BibitemOpen
  \bibfield  {author} {\bibinfo {author} {\bibfnamefont {A.~W.~P.}\
  \bibnamefont {Fitzpatrick}}, \bibinfo {author} {\bibfnamefont {S.~T.}\
  \bibnamefont {Park}}, \ and\ \bibinfo {author} {\bibfnamefont {A.~H.}\
  \bibnamefont {Zewail}},\ }\href@noop {} {\bibfield  {journal} {\bibinfo
  {journal} {PNAS}\ }\textbf {\bibinfo {volume} {110}},\ \bibinfo {pages}
  {10976} (\bibinfo {year} {2013}{\natexlab{a}})}\BibitemShut {NoStop}%
\bibitem [{\citenamefont {Fitzpatrick}\ \emph
  {et~al.}(2013{\natexlab{b}})\citenamefont {Fitzpatrick}, \citenamefont
  {Lorenz}, \citenamefont {Vanacore},\ and\ \citenamefont
  {Zewail}}]{Anthony-W.-P.-Fitzpatrick20134D-Cryo-Electro}%
  \BibitemOpen
  \bibfield  {author} {\bibinfo {author} {\bibfnamefont {A.~W.~P.}\
  \bibnamefont {Fitzpatrick}}, \bibinfo {author} {\bibfnamefont {U.~J.}\
  \bibnamefont {Lorenz}}, \bibinfo {author} {\bibfnamefont {G.~M.}\
  \bibnamefont {Vanacore}}, \ and\ \bibinfo {author} {\bibfnamefont {A.~H.}\
  \bibnamefont {Zewail}},\ }\href@noop {} {\bibfield  {journal} {\bibinfo
  {journal} {J. Am. Chem. Soc.}\ }\textbf {\bibinfo {volume} {135}},\ \bibinfo
  {pages} {19123−} (\bibinfo {year} {2013}{\natexlab{b}})}\BibitemShut
  {NoStop}%
\bibitem [{\citenamefont {Morimoto}\ and\ \citenamefont
  {Baum}(2018)}]{Baum2017NatPhys}%
  \BibitemOpen
  \bibfield  {author} {\bibinfo {author} {\bibfnamefont {Y.}~\bibnamefont
  {Morimoto}}\ and\ \bibinfo {author} {\bibfnamefont {P.}~\bibnamefont
  {Baum}},\ }\href@noop {} {\bibfield  {journal} {\bibinfo  {journal} {Nat.
  Phys.}\ }\textbf {\bibinfo {volume} {14}},\ \bibinfo {pages} {252} (\bibinfo
  {year} {2018})}\BibitemShut {NoStop}%
\bibitem [{\citenamefont {Wu}\ \emph {et~al.}(2010)\citenamefont {Wu},
  \citenamefont {Meyer-ter Vehn}, \citenamefont {Fern\'andez},\ and\
  \citenamefont {Hegelich}}]{PhysRevLett.104.234801}%
  \BibitemOpen
  \bibfield  {author} {\bibinfo {author} {\bibfnamefont {H.-C.}\ \bibnamefont
  {Wu}}, \bibinfo {author} {\bibfnamefont {J.}~\bibnamefont {Meyer-ter Vehn}},
  \bibinfo {author} {\bibfnamefont {J.}~\bibnamefont {Fern\'andez}}, \ and\
  \bibinfo {author} {\bibfnamefont {B.~M.}\ \bibnamefont {Hegelich}},\ }\href
  {\doibase 10.1103/PhysRevLett.104.234801} {\bibfield  {journal} {\bibinfo
  {journal} {Phys. Rev. Lett.}\ }\textbf {\bibinfo {volume} {104}},\ \bibinfo
  {pages} {234801} (\bibinfo {year} {2010})}\BibitemShut {NoStop}%
\bibitem [{\citenamefont {Wu}\ \emph {et~al.}(2011)\citenamefont {Wu},
  \citenamefont {Meyer-ter Vehn}, \citenamefont {Hegelich},\ and\ \citenamefont
  {Fern\'andez}}]{PhysRevSTAB.14.070702}%
  \BibitemOpen
  \bibfield  {author} {\bibinfo {author} {\bibfnamefont {H.-C.}\ \bibnamefont
  {Wu}}, \bibinfo {author} {\bibfnamefont {J.}~\bibnamefont {Meyer-ter Vehn}},
  \bibinfo {author} {\bibfnamefont {B.~M.}\ \bibnamefont {Hegelich}}, \ and\
  \bibinfo {author} {\bibfnamefont {J.~C.}\ \bibnamefont {Fern\'andez}},\
  }\href {\doibase 10.1103/PhysRevSTAB.14.070702} {\bibfield  {journal}
  {\bibinfo  {journal} {Phys. Rev. ST Accel. Beams}\ }\textbf {\bibinfo
  {volume} {14}},\ \bibinfo {pages} {070702} (\bibinfo {year}
  {2011})}\BibitemShut {NoStop}%
\bibitem [{\citenamefont {Kiefer}\ \emph {et~al.}(2013)\citenamefont {Kiefer}
  \emph {et~al.}}]{Kiefer_rel_elec_mirrors}%
  \BibitemOpen
  \bibfield  {author} {\bibinfo {author} {\bibfnamefont {D.}~\bibnamefont
  {Kiefer}} \emph {et~al.},\ }\href {\doibase 10.1038/ncomms2775} {\bibfield
  {journal} {\bibinfo  {journal} {Nat. Commun.}\ }\textbf {\bibinfo {volume}
  {4}},\ \bibinfo {pages} {1763} (\bibinfo {year} {2013})}\BibitemShut
  {NoStop}%
\bibitem [{\citenamefont {Hu}\ and\ \citenamefont
  {Wu}(2017)}]{PhysRevLett.119.254801}%
  \BibitemOpen
  \bibfield  {author} {\bibinfo {author} {\bibfnamefont {K.}~\bibnamefont
  {Hu}}\ and\ \bibinfo {author} {\bibfnamefont {H.-C.}\ \bibnamefont {Wu}},\
  }\href {\doibase 10.1103/PhysRevLett.119.254801} {\bibfield  {journal}
  {\bibinfo  {journal} {Phys. Rev. Lett.}\ }\textbf {\bibinfo {volume} {119}},\
  \bibinfo {pages} {254801} (\bibinfo {year} {2017})}\BibitemShut {NoStop}%
\bibitem [{\citenamefont {England}\ \emph {et~al.}(2014)\citenamefont {England}
  \emph {et~al.}}]{RJEngland_DLA}%
  \BibitemOpen
  \bibfield  {author} {\bibinfo {author} {\bibfnamefont {R.~J.}\ \bibnamefont
  {England}} \emph {et~al.},\ }\href@noop {} {\bibfield  {journal} {\bibinfo
  {journal} {Rev. Mod. Phys.}\ }\textbf {\bibinfo {volume} {86}},\ \bibinfo
  {pages} {1337} (\bibinfo {year} {2014})}\BibitemShut {NoStop}%
\bibitem [{\citenamefont {Ferarri}\ \emph {et~al.}(2018)\citenamefont {Ferarri}
  \emph {et~al.}}]{FerrariE_ACHIP}%
  \BibitemOpen
  \bibfield  {author} {\bibinfo {author} {\bibfnamefont {E.}~\bibnamefont
  {Ferarri}} \emph {et~al.},\ }\href@noop {} {\bibfield  {journal} {\bibinfo
  {journal} {Nucl. Instrum. Methods. Phys. Res. A}\ }\textbf {\bibinfo {volume}
  {907}},\ \bibinfo {pages} {244} (\bibinfo {year} {2018})}\BibitemShut
  {NoStop}%
\bibitem [{\citenamefont {Wang}\ and\ \citenamefont
  {Gedik}(2012)}]{WangGedikIEEE2012}%
  \BibitemOpen
  \bibfield  {author} {\bibinfo {author} {\bibfnamefont {Y.}~\bibnamefont
  {Wang}}\ and\ \bibinfo {author} {\bibfnamefont {N.}~\bibnamefont {Gedik}},\
  }\href {\doibase 10.1109/JSTQE.2011.2112339} {\bibfield  {journal} {\bibinfo
  {journal} {IEEE Journal of Selected Topics in Quantum Electronics}\ }\textbf
  {\bibinfo {volume} {18}},\ \bibinfo {pages} {140} (\bibinfo {year}
  {2012})}\BibitemShut {NoStop}%
\bibitem [{\citenamefont {Gao}\ \emph {et~al.}(2012)\citenamefont {Gao} \emph
  {et~al.}}]{Gao12}%
  \BibitemOpen
  \bibfield  {author} {\bibinfo {author} {\bibfnamefont {M.}~\bibnamefont
  {Gao}} \emph {et~al.},\ }\href {\doibase 10.1364/OE.20.012048} {\bibfield
  {journal} {\bibinfo  {journal} {Opt. Express}\ }\textbf {\bibinfo {volume}
  {20}},\ \bibinfo {pages} {12048} (\bibinfo {year} {2012})}\BibitemShut
  {NoStop}%
\bibitem [{\citenamefont {van Oudheusden}\ \emph {et~al.}(2010)\citenamefont
  {van Oudheusden} \emph {et~al.}}]{vanOudheusden2010PRL}%
  \BibitemOpen
  \bibfield  {author} {\bibinfo {author} {\bibfnamefont {T.}~\bibnamefont {van
  Oudheusden}} \emph {et~al.},\ }\href {\doibase
  10.1103/PhysRevLett.105.264801} {\bibfield  {journal} {\bibinfo  {journal}
  {Phys. Rev. Lett.}\ }\textbf {\bibinfo {volume} {105}},\ \bibinfo {pages}
  {264801} (\bibinfo {year} {2010})}\BibitemShut {NoStop}%
\bibitem [{\citenamefont {Chatelain}\ \emph {et~al.}(2012)\citenamefont
  {Chatelain}, \citenamefont {Morrison}, \citenamefont {Godbout},\ and\
  \citenamefont {Siwick}}]{Chatelain2012}%
  \BibitemOpen
  \bibfield  {author} {\bibinfo {author} {\bibfnamefont {R.~P.}\ \bibnamefont
  {Chatelain}}, \bibinfo {author} {\bibfnamefont {V.~R.}\ \bibnamefont
  {Morrison}}, \bibinfo {author} {\bibfnamefont {C.}~\bibnamefont {Godbout}}, \
  and\ \bibinfo {author} {\bibfnamefont {B.~J.}\ \bibnamefont {Siwick}},\
  }\href {\doibase 10.1063/1.4747155} {\bibfield  {journal} {\bibinfo
  {journal} {Appl. Phys. Lett.}\ }\textbf {\bibinfo {volume} {101}},\ \bibinfo
  {pages} {081901} (\bibinfo {year} {2012})}\BibitemShut {NoStop}%
\bibitem [{\citenamefont {Kassier}\ \emph {et~al.}(2012)\citenamefont {Kassier}
  \emph {et~al.}}]{Kassier2012}%
  \BibitemOpen
  \bibfield  {author} {\bibinfo {author} {\bibfnamefont {G.~H.}\ \bibnamefont
  {Kassier}} \emph {et~al.},\ }\href {\doibase 10.1007/s00340-012-5207-2}
  {\bibfield  {journal} {\bibinfo  {journal} {Appl. Phys. B}\ }\textbf
  {\bibinfo {volume} {109}},\ \bibinfo {pages} {249} (\bibinfo {year}
  {2012})}\BibitemShut {NoStop}%
\bibitem [{\citenamefont {Gliserin}\ \emph {et~al.}(2015)\citenamefont
  {Gliserin}, \citenamefont {Walbran}, \citenamefont {Krausz},\ and\
  \citenamefont {Baum}}]{Gliserin2015Sub-phonon-peri}%
  \BibitemOpen
  \bibfield  {author} {\bibinfo {author} {\bibfnamefont {A.}~\bibnamefont
  {Gliserin}}, \bibinfo {author} {\bibfnamefont {M.}~\bibnamefont {Walbran}},
  \bibinfo {author} {\bibfnamefont {F.}~\bibnamefont {Krausz}}, \ and\ \bibinfo
  {author} {\bibfnamefont {P.}~\bibnamefont {Baum}},\ }\href@noop {} {\bibfield
   {journal} {\bibinfo  {journal} {Nat. Commun.}\ }\textbf {\bibinfo {volume}
  {6}},\ \bibinfo {pages} {8723} (\bibinfo {year} {2015})}\BibitemShut
  {NoStop}%
\bibitem [{\citenamefont {Baum}\ and\ \citenamefont
  {Zewail}(2007)}]{Baum2007Attosecond-elec}%
  \BibitemOpen
  \bibfield  {author} {\bibinfo {author} {\bibfnamefont {P.}~\bibnamefont
  {Baum}}\ and\ \bibinfo {author} {\bibfnamefont {A.~H.}\ \bibnamefont
  {Zewail}},\ }\href@noop {} {\bibfield  {journal} {\bibinfo  {journal} {PNAS}\
  }\textbf {\bibinfo {volume} {104}},\ \bibinfo {pages} {18409} (\bibinfo
  {year} {2007})}\BibitemShut {NoStop}%
\bibitem [{\citenamefont {Hilbert}\ \emph {et~al.}(2009)\citenamefont
  {Hilbert}, \citenamefont {Uiterwaal}, \citenamefont {Barwick}, \citenamefont
  {Batelaan},\ and\ \citenamefont {Zewail}}]{Hilbert2009Temporal-lenses}%
  \BibitemOpen
  \bibfield  {author} {\bibinfo {author} {\bibfnamefont {S.~A.}\ \bibnamefont
  {Hilbert}}, \bibinfo {author} {\bibfnamefont {C.}~\bibnamefont {Uiterwaal}},
  \bibinfo {author} {\bibfnamefont {B.}~\bibnamefont {Barwick}}, \bibinfo
  {author} {\bibfnamefont {H.}~\bibnamefont {Batelaan}}, \ and\ \bibinfo
  {author} {\bibfnamefont {A.~H.}\ \bibnamefont {Zewail}},\ }\href@noop {}
  {\bibfield  {journal} {\bibinfo  {journal} {PNAS}\ }\textbf {\bibinfo
  {volume} {106}},\ \bibinfo {pages} {10558} (\bibinfo {year}
  {2009})}\BibitemShut {NoStop}%
\bibitem [{\citenamefont {Wong}\ \emph {et~al.}(2015)\citenamefont {Wong},
  \citenamefont {Freelon}, \citenamefont {Rohwer}, \citenamefont {Gedik},\ and\
  \citenamefont {Johnson}}]{WLJ2015NJP}%
  \BibitemOpen
  \bibfield  {author} {\bibinfo {author} {\bibfnamefont {L.~J.}\ \bibnamefont
  {Wong}}, \bibinfo {author} {\bibfnamefont {B.}~\bibnamefont {Freelon}},
  \bibinfo {author} {\bibfnamefont {T.}~\bibnamefont {Rohwer}}, \bibinfo
  {author} {\bibfnamefont {N.}~\bibnamefont {Gedik}}, \ and\ \bibinfo {author}
  {\bibfnamefont {S.~G.}\ \bibnamefont {Johnson}},\ }\href@noop {} {\bibfield
  {journal} {\bibinfo  {journal} {New J. Phys.}\ }\textbf {\bibinfo {volume}
  {17}},\ \bibinfo {pages} {013051} (\bibinfo {year} {2015})}\BibitemShut
  {NoStop}%
\bibitem [{\citenamefont {Priebe}\ \emph {et~al.}(2017)\citenamefont {Priebe}
  \emph {et~al.}}]{PriebeNatPhoton}%
  \BibitemOpen
  \bibfield  {author} {\bibinfo {author} {\bibfnamefont {K.~E.}\ \bibnamefont
  {Priebe}} \emph {et~al.},\ }\href@noop {} {\bibfield  {journal} {\bibinfo
  {journal} {Nat. Photon.}\ }\textbf {\bibinfo {volume} {11}},\ \bibinfo
  {pages} {793} (\bibinfo {year} {2017})}\BibitemShut {NoStop}%
\bibitem [{\citenamefont {Koz\'ak}\ \emph
  {et~al.}(2018{\natexlab{a}})\citenamefont {Koz\'ak}, \citenamefont
  {Eckstein}, \citenamefont {Sch\"onenberger},\ and\ \citenamefont
  {Hommelhoff}}]{KozakNatPhys2017}%
  \BibitemOpen
  \bibfield  {author} {\bibinfo {author} {\bibfnamefont {M.}~\bibnamefont
  {Koz\'ak}}, \bibinfo {author} {\bibfnamefont {T.}~\bibnamefont {Eckstein}},
  \bibinfo {author} {\bibfnamefont {N.}~\bibnamefont {Sch\"onenberger}}, \ and\
  \bibinfo {author} {\bibfnamefont {P.}~\bibnamefont {Hommelhoff}},\
  }\href@noop {} {\bibfield  {journal} {\bibinfo  {journal} {Nat. Phys.}\
  }\textbf {\bibinfo {volume} {14}},\ \bibinfo {pages} {121–125} (\bibinfo
  {year} {2018}{\natexlab{a}})}\BibitemShut {NoStop}%
\bibitem [{\citenamefont {Koz\'ak}\ \emph
  {et~al.}(2018{\natexlab{b}})\citenamefont {Koz\'ak}, \citenamefont
  {Sch\"onenberger},\ and\ \citenamefont
  {Hommelhoff}}]{PhysRevLett.120.103203}%
  \BibitemOpen
  \bibfield  {author} {\bibinfo {author} {\bibfnamefont {M.}~\bibnamefont
  {Koz\'ak}}, \bibinfo {author} {\bibfnamefont {N.}~\bibnamefont
  {Sch\"onenberger}}, \ and\ \bibinfo {author} {\bibfnamefont {P.}~\bibnamefont
  {Hommelhoff}},\ }\href {\doibase 10.1103/PhysRevLett.120.103203} {\bibfield
  {journal} {\bibinfo  {journal} {Phys. Rev. Lett.}\ }\textbf {\bibinfo
  {volume} {120}},\ \bibinfo {pages} {103203} (\bibinfo {year}
  {2018}{\natexlab{b}})}\BibitemShut {NoStop}%
\bibitem [{\citenamefont {Kealhofer}\ \emph {et~al.}(2016)\citenamefont
  {Kealhofer} \emph {et~al.}}]{KealhoferSci2016}%
  \BibitemOpen
  \bibfield  {author} {\bibinfo {author} {\bibfnamefont {C.}~\bibnamefont
  {Kealhofer}} \emph {et~al.},\ }\href@noop {} {\bibfield  {journal} {\bibinfo
  {journal} {Science}\ }\textbf {\bibinfo {volume} {352}},\ \bibinfo {pages}
  {429} (\bibinfo {year} {2016})}\BibitemShut {NoStop}%
\bibitem [{\citenamefont {Ehberger}\ \emph {et~al.}(2017)\citenamefont
  {Ehberger}, \citenamefont {Kealhofer}, \citenamefont {Krausz},\ and\
  \citenamefont {Baum}}]{EhbergerOSA2017}%
  \BibitemOpen
  \bibfield  {author} {\bibinfo {author} {\bibfnamefont {D.~P.}\ \bibnamefont
  {Ehberger}}, \bibinfo {author} {\bibfnamefont {C.}~\bibnamefont {Kealhofer}},
  \bibinfo {author} {\bibfnamefont {F.}~\bibnamefont {Krausz}}, \ and\ \bibinfo
  {author} {\bibfnamefont {P.}~\bibnamefont {Baum}},\ }in\ \href {\doibase
  10.1364/NLO.2017.NW2A.2} {\emph {\bibinfo {booktitle} {Nonlinear Optics}}}\
  (\bibinfo  {publisher} {Optical Society of America},\ \bibinfo {year}
  {2017})\ p.\ \bibinfo {pages} {NW2A.2}\BibitemShut {NoStop}%
\bibitem [{\citenamefont {Zewail}(2010)}]{Zewail187}%
  \BibitemOpen
  \bibfield  {author} {\bibinfo {author} {\bibfnamefont {A.~H.}\ \bibnamefont
  {Zewail}},\ }\href {\doibase 10.1126/science.1166135} {\bibfield  {journal}
  {\bibinfo  {journal} {Science}\ }\textbf {\bibinfo {volume} {328}},\ \bibinfo
  {pages} {187} (\bibinfo {year} {2010})}\BibitemShut {NoStop}%
\bibitem [{\citenamefont {Aidelsburger}\ \emph {et~al.}(2010)\citenamefont
  {Aidelsburger}, \citenamefont {Kirchner}, \citenamefont {Krausz},\ and\
  \citenamefont {Baum}}]{Aidelsburger2010PNAS}%
  \BibitemOpen
  \bibfield  {author} {\bibinfo {author} {\bibfnamefont {M.}~\bibnamefont
  {Aidelsburger}}, \bibinfo {author} {\bibfnamefont {F.~O.}\ \bibnamefont
  {Kirchner}}, \bibinfo {author} {\bibfnamefont {F.}~\bibnamefont {Krausz}}, \
  and\ \bibinfo {author} {\bibfnamefont {P.}~\bibnamefont {Baum}},\ }\href
  {\doibase 10.1073/pnas.1010165107} {\bibfield  {journal} {\bibinfo  {journal}
  {PNAS}\ }\textbf {\bibinfo {volume} {107}},\ \bibinfo {pages} {19714}
  (\bibinfo {year} {2010})}\BibitemShut {NoStop}%
\bibitem [{\citenamefont {Koz\'ak}(2018)}]{PhysRevA.98.013407}%
  \BibitemOpen
  \bibfield  {author} {\bibinfo {author} {\bibfnamefont {M.}~\bibnamefont
  {Koz\'ak}},\ }\href {\doibase 10.1103/PhysRevA.98.013407} {\bibfield
  {journal} {\bibinfo  {journal} {Phys. Rev. A}\ }\textbf {\bibinfo {volume}
  {98}},\ \bibinfo {pages} {013407} (\bibinfo {year} {2018})}\BibitemShut
  {NoStop}%
\bibitem [{\citenamefont {Hafizi}\ \emph {et~al.}(1997)\citenamefont {Hafizi},
  \citenamefont {Ting}, \citenamefont {Esarey}, \citenamefont {Sprangle},\ and\
  \citenamefont {Krall}}]{Hafizi1997Vacuum-beat-wav}%
  \BibitemOpen
  \bibfield  {author} {\bibinfo {author} {\bibfnamefont {B.}~\bibnamefont
  {Hafizi}}, \bibinfo {author} {\bibfnamefont {A.}~\bibnamefont {Ting}},
  \bibinfo {author} {\bibfnamefont {E.}~\bibnamefont {Esarey}}, \bibinfo
  {author} {\bibfnamefont {P.}~\bibnamefont {Sprangle}}, \ and\ \bibinfo
  {author} {\bibfnamefont {J.}~\bibnamefont {Krall}},\ }\href@noop {}
  {\bibfield  {journal} {\bibinfo  {journal} {Phys. Rev. E}\ }\textbf {\bibinfo
  {volume} {55}},\ \bibinfo {pages} {5924 } (\bibinfo {year}
  {1997})}\BibitemShut {NoStop}%
\bibitem [{\citenamefont {Koz\'ak}(2015)}]{Kozak2015Electron-accele}%
  \BibitemOpen
  \bibfield  {author} {\bibinfo {author} {\bibfnamefont {M.}~\bibnamefont
  {Koz\'ak}},\ }\href {http://stacks.iop.org/0953-4075/48/i=19/a=195601}
  {\bibfield  {journal} {\bibinfo  {journal} {J. Phys. B: At. Mol. Opt. Phys.}\
  }\textbf {\bibinfo {volume} {48}},\ \bibinfo {pages} {195601} (\bibinfo
  {year} {2015})}\BibitemShut {NoStop}%
\bibitem [{\citenamefont {Esarey}\ \emph {et~al.}(1995)\citenamefont {Esarey},
  \citenamefont {Sprangle},\ and\ \citenamefont {Krall}}]{Esarey1995PRE}%
  \BibitemOpen
  \bibfield  {author} {\bibinfo {author} {\bibfnamefont {E.}~\bibnamefont
  {Esarey}}, \bibinfo {author} {\bibfnamefont {P.}~\bibnamefont {Sprangle}}, \
  and\ \bibinfo {author} {\bibfnamefont {J.}~\bibnamefont {Krall}},\ }\href
  {\doibase 10.1103/PhysRevE.52.5443} {\bibfield  {journal} {\bibinfo
  {journal} {Phys. Rev. E}\ }\textbf {\bibinfo {volume} {52}},\ \bibinfo
  {pages} {5443} (\bibinfo {year} {1995})}\BibitemShut {NoStop}%
\bibitem [{\citenamefont {Maxson}\ \emph {et~al.}(2017)\citenamefont {Maxson},
  \citenamefont {Cesar}, \citenamefont {Calmasini}, \citenamefont {Ody},
  \citenamefont {Musumeci},\ and\ \citenamefont
  {Alesini}}]{Maxson2017Direct-Measurem}%
  \BibitemOpen
  \bibfield  {author} {\bibinfo {author} {\bibfnamefont {J.}~\bibnamefont
  {Maxson}}, \bibinfo {author} {\bibfnamefont {D.}~\bibnamefont {Cesar}},
  \bibinfo {author} {\bibfnamefont {G.}~\bibnamefont {Calmasini}}, \bibinfo
  {author} {\bibfnamefont {A.}~\bibnamefont {Ody}}, \bibinfo {author}
  {\bibfnamefont {P.}~\bibnamefont {Musumeci}}, \ and\ \bibinfo {author}
  {\bibfnamefont {D.}~\bibnamefont {Alesini}},\ }\href {\doibase
  10.1103/PhysRevLett.118.154802} {\bibfield  {journal} {\bibinfo  {journal}
  {Phys. Rev. Lett.}\ }\textbf {\bibinfo {volume} {118}},\ \bibinfo {pages}
  {154802} (\bibinfo {year} {2017})}\BibitemShut {NoStop}%
\bibitem [{\citenamefont {Zhu}\ \emph {et~al.}(2015)\citenamefont {Zhu} \emph
  {et~al.}}]{fs_time_res_diffraction}%
  \BibitemOpen
  \bibfield  {author} {\bibinfo {author} {\bibfnamefont {P.}~\bibnamefont
  {Zhu}} \emph {et~al.},\ }\href
  {http://stacks.iop.org/1367-2630/17/i=6/a=063004} {\bibfield  {journal}
  {\bibinfo  {journal} {New J. Phys.}\ }\textbf {\bibinfo {volume} {17}},\
  \bibinfo {pages} {063004} (\bibinfo {year} {2015})}\BibitemShut {NoStop}%
\bibitem [{\citenamefont {Weathersby}\ \emph {et~al.}(2015)\citenamefont
  {Weathersby} \emph {et~al.}}]{SLAC_MeV_rev_sci_instr}%
  \BibitemOpen
  \bibfield  {author} {\bibinfo {author} {\bibfnamefont {S.~P.}\ \bibnamefont
  {Weathersby}} \emph {et~al.},\ }\href {\doibase 10.1063/1.4926994} {\bibfield
   {journal} {\bibinfo  {journal} {Rev. Sci. Instrum.}\ }\textbf {\bibinfo
  {volume} {86}},\ \bibinfo {pages} {073702} (\bibinfo {year}
  {2015})}\BibitemShut {NoStop}%
\bibitem [{\citenamefont {Manz}\ \emph {et~al.}(2015)\citenamefont {Manz} \emph
  {et~al.}}]{C4FD00204K}%
  \BibitemOpen
  \bibfield  {author} {\bibinfo {author} {\bibfnamefont {S.}~\bibnamefont
  {Manz}} \emph {et~al.},\ }\href {\doibase 10.1039/C4FD00204K} {\bibfield
  {journal} {\bibinfo  {journal} {Faraday Discuss.}\ }\textbf {\bibinfo
  {volume} {177}},\ \bibinfo {pages} {467} (\bibinfo {year}
  {2015})}\BibitemShut {NoStop}%
\bibitem [{\citenamefont {Yeh}\ \emph {et~al.}(2007)\citenamefont {Yeh},
  \citenamefont {Hoffmann}, \citenamefont {Hebling},\ and\ \citenamefont
  {Nelson}}]{Yeh2007Generation-of-1}%
  \BibitemOpen
  \bibfield  {author} {\bibinfo {author} {\bibfnamefont {K.~L.}\ \bibnamefont
  {Yeh}}, \bibinfo {author} {\bibfnamefont {M.~C.}\ \bibnamefont {Hoffmann}},
  \bibinfo {author} {\bibfnamefont {J.}~\bibnamefont {Hebling}}, \ and\
  \bibinfo {author} {\bibfnamefont {K.~A.}\ \bibnamefont {Nelson}},\ }\href
  {\doibase 10.1063/1.2734374} {\bibfield  {journal} {\bibinfo  {journal}
  {Appl. Phys. Lett.}\ }\textbf {\bibinfo {volume} {90}},\ \bibinfo {pages}
  {171121} (\bibinfo {year} {2007})}\BibitemShut {NoStop}%
\bibitem [{\citenamefont {Hirori}\ \emph {et~al.}(2011)\citenamefont {Hirori},
  \citenamefont {Doi}, \citenamefont {Blanchard},\ and\ \citenamefont
  {Tanaka}}]{single_cycle_1THz}%
  \BibitemOpen
  \bibfield  {author} {\bibinfo {author} {\bibfnamefont {H.}~\bibnamefont
  {Hirori}}, \bibinfo {author} {\bibfnamefont {A.}~\bibnamefont {Doi}},
  \bibinfo {author} {\bibfnamefont {F.}~\bibnamefont {Blanchard}}, \ and\
  \bibinfo {author} {\bibfnamefont {K.}~\bibnamefont {Tanaka}},\ }\href
  {\doibase 10.1063/1.3560062} {\bibfield  {journal} {\bibinfo  {journal}
  {Appl. Phys. Lett.}\ }\textbf {\bibinfo {volume} {98}},\ \bibinfo {pages}
  {091106} (\bibinfo {year} {2011})}\BibitemShut {NoStop}%
\bibitem [{\citenamefont {Hauri}\ \emph {et~al.}(2011)\citenamefont {Hauri},
  \citenamefont {Ruchert}, \citenamefont {Vicario},\ and\ \citenamefont
  {Ardana}}]{OR_organic_crystals}%
  \BibitemOpen
  \bibfield  {author} {\bibinfo {author} {\bibfnamefont {C.~P.}\ \bibnamefont
  {Hauri}}, \bibinfo {author} {\bibfnamefont {C.}~\bibnamefont {Ruchert}},
  \bibinfo {author} {\bibfnamefont {C.}~\bibnamefont {Vicario}}, \ and\
  \bibinfo {author} {\bibfnamefont {F.}~\bibnamefont {Ardana}},\ }\href
  {\doibase 10.1063/1.3655331} {\bibfield  {journal} {\bibinfo  {journal}
  {Appl. Phys. Lett.}\ }\textbf {\bibinfo {volume} {99}},\ \bibinfo {pages}
  {161116} (\bibinfo {year} {2011})}\BibitemShut {NoStop}%
\bibitem [{\citenamefont {Sell}\ \emph {et~al.}(2008)\citenamefont {Sell},
  \citenamefont {Leitenstorfer},\ and\ \citenamefont {Huber}}]{THz_DFG}%
  \BibitemOpen
  \bibfield  {author} {\bibinfo {author} {\bibfnamefont {A.}~\bibnamefont
  {Sell}}, \bibinfo {author} {\bibfnamefont {A.}~\bibnamefont {Leitenstorfer}},
  \ and\ \bibinfo {author} {\bibfnamefont {R.}~\bibnamefont {Huber}},\
  }\href@noop {} {\bibfield  {journal} {\bibinfo  {journal} {Opt. Lett.}\
  }\textbf {\bibinfo {volume} {33}},\ \bibinfo {pages} {2767} (\bibinfo {year}
  {2008})}\BibitemShut {NoStop}%
\bibitem [{\citenamefont {Huang}\ \emph {et~al.}(2013)\citenamefont {Huang}
  \emph {et~al.}}]{HuangOptLett2013}%
  \BibitemOpen
  \bibfield  {author} {\bibinfo {author} {\bibfnamefont {S.~W.}\ \bibnamefont
  {Huang}} \emph {et~al.},\ }\href {\doibase 10.1364/OL.38.000796} {\bibfield
  {journal} {\bibinfo  {journal} {Opt. Lett.}\ }\textbf {\bibinfo {volume}
  {38}},\ \bibinfo {pages} {796} (\bibinfo {year} {2013})}\BibitemShut
  {NoStop}%
\bibitem [{\citenamefont {Dhillon}\ \emph {et~al.}(2017)\citenamefont {Dhillon}
  \emph {et~al.}}]{Dhillon2017a}%
  \BibitemOpen
  \bibfield  {author} {\bibinfo {author} {\bibfnamefont {S.}~\bibnamefont
  {Dhillon}} \emph {et~al.},\ }\href
  {http://stacks.iop.org/0022-3727/50/i=4/a=043001} {\bibfield  {journal}
  {\bibinfo  {journal} {J. Phys. D: Appl. Phys.}\ }\textbf {\bibinfo {volume}
  {50}},\ \bibinfo {pages} {043001} (\bibinfo {year} {2017})}\BibitemShut
  {NoStop}%
\bibitem [{\citenamefont {Fulop}\ \emph {et~al.}(2010)\citenamefont {Fulop},
  \citenamefont {Palfalvi}, \citenamefont {Almasi},\ and\ \citenamefont
  {Hebling}}]{Fulop_THz_OR}%
  \BibitemOpen
  \bibfield  {author} {\bibinfo {author} {\bibfnamefont {J.~A.}\ \bibnamefont
  {Fulop}}, \bibinfo {author} {\bibfnamefont {L.}~\bibnamefont {Palfalvi}},
  \bibinfo {author} {\bibfnamefont {G.}~\bibnamefont {Almasi}}, \ and\ \bibinfo
  {author} {\bibfnamefont {J.}~\bibnamefont {Hebling}},\ }\href@noop {}
  {\bibfield  {journal} {\bibinfo  {journal} {Opt. Express}\ }\textbf {\bibinfo
  {volume} {18}},\ \bibinfo {pages} {12311} (\bibinfo {year}
  {2010})}\BibitemShut {NoStop}%
\bibitem [{\citenamefont {Fulop}\ \emph {et~al.}(2011)\citenamefont {Fulop},
  \citenamefont {Palfalvi}, \citenamefont {Hoffmann},\ and\ \citenamefont
  {Hebling}}]{Fulop_mJ_THz}%
  \BibitemOpen
  \bibfield  {author} {\bibinfo {author} {\bibfnamefont {J.}~\bibnamefont
  {Fulop}}, \bibinfo {author} {\bibfnamefont {L.}~\bibnamefont {Palfalvi}},
  \bibinfo {author} {\bibfnamefont {M.~C.}\ \bibnamefont {Hoffmann}}, \ and\
  \bibinfo {author} {\bibfnamefont {J.}~\bibnamefont {Hebling}},\ }\href@noop
  {} {\bibfield  {journal} {\bibinfo  {journal} {Opt. Express}\ }\textbf
  {\bibinfo {volume} {19}},\ \bibinfo {pages} {15090} (\bibinfo {year}
  {2011})}\BibitemShut {NoStop}%
\bibitem [{\citenamefont {Fulop}\ \emph {et~al.}(2014)\citenamefont {Fulop}
  \emph {et~al.}}]{THz_0.4mJ}%
  \BibitemOpen
  \bibfield  {author} {\bibinfo {author} {\bibfnamefont {J.~A.}\ \bibnamefont
  {Fulop}} \emph {et~al.},\ }\href@noop {} {\bibfield  {journal} {\bibinfo
  {journal} {Opt. Express}\ }\textbf {\bibinfo {volume} {22}},\ \bibinfo
  {pages} {20155} (\bibinfo {year} {2014})}\BibitemShut {NoStop}%
\bibitem [{\citenamefont {Vicario}\ \emph {et~al.}(2014)\citenamefont {Vicario}
  \emph {et~al.}}]{THz_0.9mJ}%
  \BibitemOpen
  \bibfield  {author} {\bibinfo {author} {\bibfnamefont {C.}~\bibnamefont
  {Vicario}} \emph {et~al.},\ }\href@noop {} {\bibfield  {journal} {\bibinfo
  {journal} {Opt. Lett.}\ }\textbf {\bibinfo {volume} {39}},\ \bibinfo {pages}
  {6632} (\bibinfo {year} {2014})}\BibitemShut {NoStop}%
\bibitem [{\citenamefont {Hastings}\ \emph {et~al.}(2006)\citenamefont
  {Hastings} \emph {et~al.}}]{HastingsApplPhysLett2006}%
  \BibitemOpen
  \bibfield  {author} {\bibinfo {author} {\bibfnamefont {J.~B.}\ \bibnamefont
  {Hastings}} \emph {et~al.},\ }\href {\doibase 10.1063/1.2372697} {\bibfield
  {journal} {\bibinfo  {journal} {Applied Physics Letters}\ }\textbf {\bibinfo
  {volume} {89}},\ \bibinfo {pages} {184109} (\bibinfo {year}
  {2006})}\BibitemShut {NoStop}%
\bibitem [{\citenamefont {Musumeci}\ \emph
  {et~al.}(2010{\natexlab{b}})\citenamefont {Musumeci}, \citenamefont {Moody},
  \citenamefont {Scoby}, \citenamefont {Gutierrez},\ and\ \citenamefont
  {Westfall}}]{MusumeciApplPhysLett2010}%
  \BibitemOpen
  \bibfield  {author} {\bibinfo {author} {\bibfnamefont {P.}~\bibnamefont
  {Musumeci}}, \bibinfo {author} {\bibfnamefont {J.~T.}\ \bibnamefont {Moody}},
  \bibinfo {author} {\bibfnamefont {C.~M.}\ \bibnamefont {Scoby}}, \bibinfo
  {author} {\bibfnamefont {M.~S.}\ \bibnamefont {Gutierrez}}, \ and\ \bibinfo
  {author} {\bibfnamefont {M.}~\bibnamefont {Westfall}},\ }\href@noop {}
  {\bibfield  {journal} {\bibinfo  {journal} {Appl. Phys. Lett.}\ }\textbf
  {\bibinfo {volume} {97}},\ \bibinfo {pages} {063502} (\bibinfo {year}
  {2010}{\natexlab{b}})}\BibitemShut {NoStop}%
\bibitem [{\citenamefont {Zeitler}\ \emph {et~al.}(2015)\citenamefont
  {Zeitler}, \citenamefont {Floettmann},\ and\ \citenamefont
  {Gr\"uner}}]{PhysRevSTAB.18.120102}%
  \BibitemOpen
  \bibfield  {author} {\bibinfo {author} {\bibfnamefont {B.}~\bibnamefont
  {Zeitler}}, \bibinfo {author} {\bibfnamefont {K.}~\bibnamefont {Floettmann}},
  \ and\ \bibinfo {author} {\bibfnamefont {F.}~\bibnamefont {Gr\"uner}},\
  }\href {\doibase 10.1103/PhysRevSTAB.18.120102} {\bibfield  {journal}
  {\bibinfo  {journal} {Phys. Rev. ST Accel. Beams}\ }\textbf {\bibinfo
  {volume} {18}},\ \bibinfo {pages} {120102} (\bibinfo {year}
  {2015})}\BibitemShut {NoStop}%
\bibitem [{\citenamefont {Kimura}\ \emph {et~al.}(2004)\citenamefont {Kimura}
  \emph {et~al.}}]{PRL_hightrapping_efficiency}%
  \BibitemOpen
  \bibfield  {author} {\bibinfo {author} {\bibfnamefont {W.~D.}\ \bibnamefont
  {Kimura}} \emph {et~al.},\ }\href@noop {} {\bibfield  {journal} {\bibinfo
  {journal} {Phys. Rev. Lett.}\ }\textbf {\bibinfo {volume} {92}},\ \bibinfo
  {pages} {1} (\bibinfo {year} {2004})}\BibitemShut {NoStop}%
\bibitem [{\citenamefont {Tibai}\ \emph {et~al.}(2014)\citenamefont {Tibai},
  \citenamefont {T\'oth}, \citenamefont {Mechler}, \citenamefont {F\"ul\"op},
  \citenamefont {Alm\'asi},\ and\ \citenamefont
  {Hebling}}]{PRL_single_cycle_XUV}%
  \BibitemOpen
  \bibfield  {author} {\bibinfo {author} {\bibfnamefont {Z.}~\bibnamefont
  {Tibai}}, \bibinfo {author} {\bibfnamefont {G.}~\bibnamefont {T\'oth}},
  \bibinfo {author} {\bibfnamefont {M.~I.}\ \bibnamefont {Mechler}}, \bibinfo
  {author} {\bibfnamefont {J.~A.}\ \bibnamefont {F\"ul\"op}}, \bibinfo {author}
  {\bibfnamefont {G.}~\bibnamefont {Alm\'asi}}, \ and\ \bibinfo {author}
  {\bibfnamefont {J.}~\bibnamefont {Hebling}},\ }\href {\doibase
  10.1103/PhysRevLett.113.104801} {\bibfield  {journal} {\bibinfo  {journal}
  {Phys. Rev. Lett.}\ }\textbf {\bibinfo {volume} {113}},\ \bibinfo {pages}
  {104801} (\bibinfo {year} {2014})}\BibitemShut {NoStop}%
\bibitem [{\citenamefont {Karpowicz}\ \emph {et~al.}(2008)\citenamefont
  {Karpowicz} \emph {et~al.}}]{THz_plasma_ionization}%
  \BibitemOpen
  \bibfield  {author} {\bibinfo {author} {\bibfnamefont {N.}~\bibnamefont
  {Karpowicz}} \emph {et~al.},\ }\href {\doibase 10.1063/1.2828709} {\bibfield
  {journal} {\bibinfo  {journal} {Appl. Phys. Lett.}\ }\textbf {\bibinfo
  {volume} {92}},\ \bibinfo {pages} {011131} (\bibinfo {year}
  {2008})}\BibitemShut {NoStop}%
\bibitem [{\citenamefont {Sergeeva}\ \emph {et~al.}(2017)\citenamefont
  {Sergeeva}, \citenamefont {Potylitsyn}, \citenamefont {Tishchenko},\ and\
  \citenamefont {Strikhanov}}]{Sergeeva2017Smith-Purcell-r}%
  \BibitemOpen
  \bibfield  {author} {\bibinfo {author} {\bibfnamefont {D.~Y.}\ \bibnamefont
  {Sergeeva}}, \bibinfo {author} {\bibfnamefont {A.~P.}\ \bibnamefont
  {Potylitsyn}}, \bibinfo {author} {\bibfnamefont {A.~A.}\ \bibnamefont
  {Tishchenko}}, \ and\ \bibinfo {author} {\bibfnamefont {M.~N.}\ \bibnamefont
  {Strikhanov}},\ }\href@noop {} {\bibfield  {journal} {\bibinfo  {journal}
  {Opt. Express}\ }\textbf {\bibinfo {volume} {25}},\ \bibinfo {pages} {26310}
  (\bibinfo {year} {2017})}\BibitemShut {NoStop}%
\bibitem [{\citenamefont {Zhang}\ \emph {et~al.}(2017)\citenamefont {Zhang}
  \emph {et~al.}}]{Zhang2017Transition-radi}%
  \BibitemOpen
  \bibfield  {author} {\bibinfo {author} {\bibfnamefont {K.}~\bibnamefont
  {Zhang}} \emph {et~al.},\ }\href@noop {} {\bibfield  {journal} {\bibinfo
  {journal} {Opt. Express}\ }\textbf {\bibinfo {volume} {25}},\ \bibinfo
  {pages} {20477} (\bibinfo {year} {2017})}\BibitemShut {NoStop}%
\bibitem [{\citenamefont {Wong}\ \emph {et~al.}(2016)\citenamefont {Wong},
  \citenamefont {Kaminer}, \citenamefont {Ilic}, \citenamefont {Joannopoulos},\
  and\ \citenamefont {Soljacic}}]{WLJ_nat_photon_2016}%
  \BibitemOpen
  \bibfield  {author} {\bibinfo {author} {\bibfnamefont {L.~J.}\ \bibnamefont
  {Wong}}, \bibinfo {author} {\bibfnamefont {I.}~\bibnamefont {Kaminer}},
  \bibinfo {author} {\bibfnamefont {O.}~\bibnamefont {Ilic}}, \bibinfo {author}
  {\bibfnamefont {J.~D.}\ \bibnamefont {Joannopoulos}}, \ and\ \bibinfo
  {author} {\bibfnamefont {M.}~\bibnamefont {Soljacic}},\ }\href@noop {}
  {\bibfield  {journal} {\bibinfo  {journal} {Nat. Photon.}\ }\textbf {\bibinfo
  {volume} {10}},\ \bibinfo {pages} {46} (\bibinfo {year} {2016})}\BibitemShut
  {NoStop}%
\bibitem [{\citenamefont {Rosolen}\ \emph {et~al.}(2018)\citenamefont
  {Rosolen}, \citenamefont {Wong}, \citenamefont {Rivera}, \citenamefont
  {Maes}, \citenamefont {Soljacic},\ and\ \citenamefont
  {Kaminer}}]{Rosolen_LightSci_2018}%
  \BibitemOpen
  \bibfield  {author} {\bibinfo {author} {\bibfnamefont {G.}~\bibnamefont
  {Rosolen}}, \bibinfo {author} {\bibfnamefont {L.~J.}\ \bibnamefont {Wong}},
  \bibinfo {author} {\bibfnamefont {N.}~\bibnamefont {Rivera}}, \bibinfo
  {author} {\bibfnamefont {B.}~\bibnamefont {Maes}}, \bibinfo {author}
  {\bibfnamefont {M.}~\bibnamefont {Soljacic}}, \ and\ \bibinfo {author}
  {\bibfnamefont {I.}~\bibnamefont {Kaminer}},\ }\href@noop {} {\bibfield
  {journal} {\bibinfo  {journal} {Light Sci. Appl.}\ }\textbf {\bibinfo
  {volume} {7}},\ \bibinfo {pages} {64} (\bibinfo {year} {2018})}\BibitemShut
  {NoStop}%
\end{thebibliography}

%

\end{document}